\documentclass[12pt,a4paper]{article}
\pdfoutput=1
 \usepackage{jheppub}
 \usepackage{amsmath}
 \usepackage{amsfonts}
 \usepackage{mathtools}
 \usepackage{amssymb}
 \usepackage{caption}
 \usepackage{subcaption}
 \usepackage{slashed}
\usepackage{float}
\usepackage{graphicx}
\graphicspath{ {project2 tex/} }

\def\al{\alpha}

\newcommand{\be} {\begin{equation}}
\newcommand{\ee} {\end{equation}}
\newcommand{\bea} {\begin{eqnarray}}
\newcommand{\eea} {\end{eqnarray}}
\newcommand{\ba} {\begin{array}}
\newcommand{\ea} {\end{array}}
\newcommand{\nn} {\nonumber}

 \title{Dyon degeneracies from Mathieu moonshine}
 \author{Aradhita Chattopadhyaya, Justin R. David}
\affiliation{Centre for High Energy Physics, Indian Institute of Science,\\
C. V. Raman Avenue, Bangalore 560012, India.}
\emailAdd{aradhita, justin@cts.iisc.ernet.in}

\abstract{
We  construct the Siegel modular forms associated with 
the theta lift of  twisted elliptic genera of $K3$ orbifolded with   $g'$ 
corresponding to the
conjugacy classes of the Mathieu group $M_{24}$. We complete the construction 
for all the  classes  which belong to $M_{23} \subset M_{24}$ and 
two other classes outside the subgroup  $M_{23}$. 
For this purpose we provide the explicit expressions for all the twisted elliptic genera 
in all the sectors of these classes. 
 We show that the Siegel modular forms 
 satisfy the required properties for them to be generating functions 
of $1/4$ BPS dyons of type II string theories compactified on  $K3\times T^2$  and
orbifolded
by $g'$ which acts as a $\mathbb{Z}_N$ automorphism on $K3$ together with a
$1/N$ shift on a circle of $T^2$. 
In particular the inverse of  these Siegel modular forms admit 
a Fourier  expansion with integer coefficients together with the right sign
as predicted from black hole physics.  Our analysis completes the construction 
of the  partition function for dyons  as well as the twisted elliptic genera for all the 
$7$ CHL compactifications. 
}

\begin{document}
\maketitle
\flushbottom

\section{Introduction}

The partition function of $1/4$ BPS dyons for ${\cal N}=4$ string compactifications
have been studied extensively.  Starting from the original proposal \cite{Dijkgraaf:1996it}
for the degeneracy of dyons in heterotic string theory on 
$T^6$ and the study of its asymptotic property \cite{LopesCardoso:2004law}, 
  it has been generalized to certain  CHL compactifications \cite{Jatkar:2005bh}. 
The degeneracy of dyons can be obtained from the Fourier coefficients of  the inverse 
of an appropriate 
 $Sp(2, \mathbb{Z})$  Siegel modular  forms or its subgroup.  
 For the case of the heterotic string on $T^6$,  it is the Igusa cusp form 
 of weight 10 which is the theta lift  or the multiplicative lift of the elliptic genus of $K3$. 
 The elliptic genus of $K3$ plays a role in the degeneracy since the counting 
 of these $1/4$ BPS states is done in the type II picture which is 
 compactified on $K3\times T^2$ \cite{David:2006yn,Shih:2005uc}. 
 For the case of CHL compactifications \cite{Chaudhuri:1995fk}  considered 
 it turns out that the Siegel modular forms are theta lifts of the twisted elliptic 
 genus of $K3$ \cite{David:2006ji,Dabholkar:2006xa}. This is because the CHL compactifications are dual 
 to  $( K3\times T^2)/\mathbb{Z}_N$ where the orbifold  acts as an 
 order  $\mathbb{Z}_N$  Nikulin's automorphism \cite{Nikulin} on  $K3$ together with 
 a $1/N$ shift on one of the circles of $T^2$ \cite{Chaudhuri:1995dj,Aspinwall:1995fw}.  The construction 
 so far has been for the case of $N=2, 3, 5, 7$ CHL models, that is $4$ out of the 
 $7$ CHL models.

 With the discovery of Mathieu moonshine in $K3$  \cite{Eguchi:2010ej}, it has been seen 
 $K3$ admits 26 twining elliptic genera corresponding to the $26$  conjugacy 
 classes of the 
 Mathieu group $M_{24}$.   Before we proceed let us define 
 the twisted elliptic genus of $K3$ by an automorphism $g'$ of order  $\mathbb{Z}_N$,  given by
 \begin{equation}
 F^{(r,s)}(\tau,z)=\frac{1}{N}{\rm Tr}_{R\, R g'^r}[(-1)^{F_{K3}+\bar{F}_{K3} }g'^s e^{2\pi i zF_{K3}}
 q^{L_0-c/24} \bar{q}^{\bar{L_0}-\bar{c}/24}].
 \end{equation}
 Here the trace is taken over the Ramond-Ramond 
 sector of the   ${\cal N}=(4, ,4 )$  superconformal 
 field theory  of $K3$ 
 with central charge $(6, 6)$ and  $F$ is the fermion number. 
 The $K3$ CFT is orbifolded by the action of $g'$,  a $\mathbb{Z}_N$ 
 automorphism. 
 The values $(r, s)$ run from $0$ to $N-1$. 
 For $g'$ belonging to the $26$ conjugacy classes of $M_{24}$ only  the twining character
 $F^{(0, 1)}$ has been constructed in  \cite{Cheng:2010pq,Eguchi:2010fg,Gaberdiel:2010ch}. 
 The names of these classes and the 
 corresponding cycle and the cycle shape are listed in tables \ref{t1} and \ref{t2}.
 The order of the corresponding automorphism in $K3$ is also listed. 
For the  $M_{24}$ conjugacy classes $pA, p =2, 3, 5, 7$,  the twisted elliptic
genera in all the sectors was given earlier in \cite{David:2006ji}.  These genera are obtained by 
orbifolding the $K3$ by $g'$ which is an order $N=2, 3, 5, 7$ automorphism. 
The Siegel modular forms which capture the degeneracy of $1/4$ BPS 
states in the ${\cal N}=4$ theories  obtained by  type II  compactified on 
the orbifold $( K3\times T^2)/\mathbb{Z}_N$ have  also been constructed in \cite{David:2006ji}. 
The most direct method of constructing these Siegel modular forms 
is thorough the theta lift of the corresponding twisted elliptic genus 
of $K3$.  For this purpose it is necessary to know  the 
Fourier expansion of the twisted elliptic 
genus in all its sectors. 
In this paper we  extend this construction of Siegel modular forms  to 
the other conjugacy classes of $M_{24}$. 
The construction  is carried out  for all classes in table \ref{t1} and for the first two classes in 
table \ref{t2}. 
 We also 
demonstrate that the inverse of these Siegel modular forms  
have the required properties to be generating functions 
of  $1/4$ BPS states of type II  string  compactified on 
orbifolds of $K3\times T^2$ by $g'$  on $K3$ corresponding to these conjugacy classes
together with a shift of $1/N$  on one of the circles of $T^2$.  
 See \cite{Sen:2007qy,Dabholkar:2012zz}  for reviews. Our main objective
 is to study the Fourier expansion of the resulting Siegel modular forms 
 and observe that their coefficients are integers as well as positive in accordance
 with the conjecture of \cite{Sen:2010mz}.  

As we have remarked the first step towards constructing the Siegel modular form
obtained as a theta lift of the twisted elliptic genus of $K3$ is 
the knowledge of the  Fourier expansion  in all the sectors $F^{(r, s)}$. 
Other than the conjugacy classes $pA, p = 2, 3, 5, 7$  only the 
twining character $F^{(0, 1)}$ is known. 
To obtain the other sectors we use the following transformation 
property of the twisted elliptic genus under modular transformation
\begin{equation} \label{fulltwist}
 F^{(r, s)} \left( \frac{a\tau + b}{ c\tau + d}, \frac{ z}{ c\tau +d}
 \right) = \exp\left(  2\pi i \frac{c z^2}{ c\tau +d}  \right) 
 F^{( cs + a r, ds + b r)} ( \tau,  z) ,
\end{equation}
with 
\begin{equation}
 a, b, c, d \in \mathbb{Z}, \qquad ad -bc =1.
\end{equation}
In (\ref{fulltwist}) the indices $cs + ar$ and $ds + br $  belong to $\mathbb{Z}$ mod $N$. 
For example the $(0, 1)$ sector on the LHS of  (\ref{fulltwist}) is 
related to the $(1, 0)$ sector its arguments is evaluated at 
$(-1/\tau, z/\tau)$. 
However this is not sufficient, since we require a Fourier expansion 
of the $(1, 0)$ sector  to construct the theta lift,  in fact we need further relations 
to express the the expansion in terms of $e^{- 2\pi i /\tau}$ in terms 
ordinary $q = e^{2\pi i \tau}$ expansions. 
We find several identities involving modular 
forms of $\Gamma_0(N)$ which will enable us to perform this explicitly in this paper. 
For $N$ prime this procedure is enough to determine all the 
sectors of the twisted elliptic genus. 
But, when $N$ is composite it is not possible to relate 
all the sectors to the $(0, 1)$ sector by modular transformation. 
The various sectors of the twisted elliptic genus  break up 
into sub-orbits under the action of modular transformations. 
For example for the class $4B$ with $N=4$ in table \ref{t1},  the sectors $F^{(0,2)}, 
F^{(2, 0)}, F^{(2,2)}$ form a sub-orbit and cannot be related to $F^{(0,1)}$. 
We determine the twisted elliptic genus in these sub-orbits 
 using its correspondence with the cycle 
shape  of $M_{24}$.  
This correspondence is sufficient to determine the complete
twisted elliptic genus for all the  classes  given in table \ref{t1}.

Among the classes in table \ref{t2},  we  construct the twisted 
elliptic genus for classes $2B$ and $3B$. 
The  cycle shape of conjugacy classes  belonging to this table  is such that 
one cannot use it to determine the twisted elliptic genus in the sub-orbits. 
For example squaring the cycle in the $2B$  class leads to the identity. 
In \cite{Gaberdiel:2013psa}, an explicit rational  CFT consisting of $6$,  $SU(2)$ WZW models
at level $1$ in  which the $2B$ orbifold 
can be performed was introduced. 
This construction enables the evaluation of the twisted elliptic genus 
in all the sectors. 
The twisted elliptic genus exhibits the following property 
which is known as `quantum symmetry'.  This essentially means that the sum 
of the twisted elliptic genus in all its sectors  vanishes. 
For example for the case of $2B$ which is an order $4$ action in $K3$ quantum symmetry 
implies the equality
\begin{equation}
 \sum_{r, s=0}^{3} F^{(r, s)}(\tau, z) =0.
\end{equation}
Using this symmetry we obtain the twisted elliptic genus 
for the $3B$ in all the sectors. 

At this point it is important to mention the the \cite{Gaberdiel:2012gf} also constructs the twisted 
elliptic genus for all the orbifolds considered in this paper as well as orbifolds by 
non-cyclic groups. A detailed comparison of our work with \cite{Gaberdiel:2012gf} will made 
in section \ref{comparision}. But we remark here that the explicit expressions 
that we derive here are not found in the main body of \cite{Gaberdiel:2012gf}.

We then construct and determine the 
weights $k$ of the Siegel modular form $\tilde\Phi_k(\rho, \sigma, v)$.
obtained from the theta lift of the twisted elliptic genus corresponding to  the conjugacy classes in table (\ref{t1})  
and the first two classes in table (\ref{t2}). 
We  study the
  factorization property on divisors as $v\rightarrow 0$. 
 This enables us to obtain the asymptotic degeneracies of $1/4$ BPS black
 holes of large charges in type II string theory compactified on the orbifold
 $(K3\times T^2)/\mathbb{Z}_N$ including the sub-leading corrections. 
 Using the analysis in \cite{David:2006ud}, we see the sub-leading corrections
 agree precisely with that obtained using the entropy function
 method including the Gauss-Bonnet term in these theories. 
We also  obtain the degeneracies of $1/2$ BPS states of these theories 
 which are either purely electric or magnetic. 
 Finally we explicitly evaluate the degeneracies of the low charge dyons in these states
  using these  Siegel modular forms by extracting out the respective Fourier coefficients. 
  This is given by the expression 
  \be \label{b6}
-B_6=-(-1)^{Q \cdot P}\int_{{\cal C}}{\rm d}\rho{\rm d}
\sigma {\rm d}v\; e^{-\pi i (N\rho Q^2+\sigma/N P^2+2v Q\cdot P)}\frac{1}{\tilde \Phi(\rho,\sigma, v)},
\ee
where ${\cal C}$ is a contour in the complex 3-plane which we will define.
$Q, P$ refer to the electric and magnetic charge of the dyons in the heterotic
frame. 
We emphasize that the evaluation of the degeneracy $-B_6$ for low charge dyons 
is possible only due to the explicit knowledge of the twisted elliptic genus 
in all its sectors. 
A subset of these 
Fourier coefficients represent single centered 
black holes. From the fact that the single centered black holes    carry zero angular momentum, 
  it is  conjectured that  the sign of $-B_6$ is positive \cite{Sen:2010mz}. 
We  verify this prediction for low charge dyons.  
All these properties of $(\tilde \Phi_k)^{-1} $ indicate that they capture 
the degeneracy of dyons in ${\cal N} =4$ theories compactified on 
orbifolds $(K3\times T^2)/\mathbb{Z}_N$  where $\mathbb{Z}_N$ 
acts as $g'$ an order $N$ automorphism in $K3$ together with 
a $1/N$ shift on one of the circles of $T^2$. 
The construction of  Siegel modular forms for the cases of composite order $4B, 6A, 8A$ 
 in table \ref{t1}
together with the earlier construction  of the Siegel modular forms for the classes 
$pA, p =2, 3, 5, 7$ completes the  study of the spectrum of $1/4$ BPS dyons in all the 
 $7$ CHL compactifications introduced in \cite{Chaudhuri:1995dj,Chaudhuri:1995fk}. 
 
 Again here we remark the construction of the modular forms $\tilde \Phi_k$ given the 
 twisted elliptic genus is quite straight forward and our method is the extension 
 of the method first introduced for cyclic orbifolds in \cite{David:2006ud}. 
 Recently this construction 
 has been extended for non-cyclic orbifolds in 
 \cite{Persson:2013xpa,Persson:2015jka,Paquette:2017gmb}. 
 However to our knowledge  that the observation of positivity of the Fourier coefficients 
 of the inverse of $\tilde \Phi_k$ which is
in agreement with the conjecture of \cite{Persson:2013xpa,Persson:2015jka,Paquette:2017gmb} 
for all the orbifolds considered in this  paper is new. 
This observation has been made possible by  the careful construction of all the twisted elliptic genera
performed in section \ref{twistgenus}. 
It is also important to note that a proof for the positivity conjecture of 
\cite{Sen:2010mz} has been made only for the case of a class of Fourier coefficients 
of the partition function $\Phi_{10}$ by \cite{Bringmann:2012zr}. 
A general proof  of the positivity conjecture for $\Phi_{10}$ as well as all the cyclic orbifolds 
considered in this paper is an open question.

 The organization  of the paper is follows:
 In section \ref{twistgenus},  we construct the twisted elliptic 
 genus for different orbifolds of $K3$ in each sector, 
 We first discuss the orbifolds of $K3$  corresponding to the classes in 
 table \ref{t1} and then move on to the classes $2B$ and $3B$ of table 
 \ref{t2}.
 In section  \ref{dydegen}, we use the twisted elliptic genus to construct the 
 Siegel modular forms that capture degeneracies of $1/4$ BPS dyons of
 type II theories compactified on $(K3\times T^2)/\mathbb{Z}_N$ where $\mathbb{Z}_N$
 acts as a order $N$ automorphism on $K3$ together with a $1/N$ shift on one 
 of the circles of $T^2$.  We show that low lying coefficients of the $1/4$ BPS index 
 are positive as expected from black hole considerations in section \ref{b66}. 
 Appendix \ref{geneta} lists various identities relating modular forms involving 
 expansions in $e^{-2\pi i /\tau}$ to $e^{2\pi i \tau}$. 
 Finally appendix \ref{frslist} lists the twisted elliptic genus for the $14A$ and the $15A$
 conjugacy class. 

\begin{table}[H]
\renewcommand{\arraystretch}{0.5}
\begin{center}
\vspace{0.5cm}
 {\scriptsize{
\begin{tabular}{|c|c|c|c|}
\hline
 & & & \\
Conjugacy Class & Order & Cycle shape & Cycle \\
\hline
 & & & \\
1A & 1 & $1^{24}$ & () \\
2A & 2  & $1^8\cdot 2^8$ & (1, 8)(2, 12)(4, 15)(5, 7)(9, 22)(11, 18)(14, 19)(23, 24) \\
3A & 3 & $1^6\cdot 3^6$ & (3, 18, 20)(4, 22, 24)(5, 19, 17)(6, 11, 8)(7, 15, 10)(9, 12, 14) \\
5A & 4 & $1^4 \cdot 5^4$ & (2, 21, 13, 16, 23)(3, 5, 15, 22, 14)(4, 12, 20, 17, 7)(9, 18, 19, 10, 24) \\
7A & 7 &  $1^3 \cdot 7^3$ & (1, 17, 5, 21, 24, 10, 6)(2, 12, 13, 9, 4, 23, 20)(3, 8, 22, 7, 18, 14, 19) \\
7A & 7 &  $1^3 \cdot 7^3$ & (1, 21, 6, 5, 10, 17, 24)(2, 9, 20, 13, 23, 12, 4)(3, 7, 19, 22, 14, 8, 18) \\
11A & 11 & $1^2 \cdot 11^2$ & (1, 3, 10, 4, 14, 15, 5, 24, 13, 17, 18)(2, 21, 23, 9, 20, 19, 6, 12, 16, 11, 22) \\
23A & 23 & $1^1 \cdot 23^1$ & (1, 7, 6, 24, 14, 4, 16, 12, 20, 9, 11, 5, 15, 10, 19, 18, 23, 17, 3, 2, 8, 22, 21)\\
23B & 23 &  $1^1 \cdot 23^1$  & (1, 4, 11, 18, 8, 6, 12, 15, 17, 21, 14, 9, 19, 2, 7, 16, 5, 23, 22, 24, 20, 10, 3) \\
 & & & \\
\hline
 & & & \\
4B & 4 & $1^4\cdot 2^2 \dot 4^4$ & (1, 17, 21, 9)(2, 13, 24, 15)(3, 23)(4, 14, 5, 8)(6, 16)(12, 18, 20, 22)\\
6A & 6 & $1^2 \cdot 2^2 \cdot 3^2 \cdot 6^2$ & (1, 8)(2, 24, 11, 12, 23, 18)(3, 20, 10)(4, 15)(5, 19, 9, 7, 14, 22)(6, 16, 13) \\
8A & 8 & $1^2 \cdot 2^1 \cdot 4^1 \cdot 8^2$ & (1, 13, 17, 24, 21, 15, 9, 2)(3, 16, 23, 6)(4, 22, 14, 12, 5, 18, 8, 20)(7, 11) \\
14A & 14 & $1^1\cdot 2^1 \cdot 7^1 \cdot 14^1$ & (1, 12, 17, 13, 5, 9, 21, 4, 24, 23, 10, 20, 6, 2)(3, 18, 8, 14, 22, 19, 7)(11, 15) \\
14B & 14 & $1^1\cdot 2^1 \cdot 7^1 \cdot 14^1$ & (1, 13, 21, 23, 6, 12, 5, 4, 10, 2, 17, 9, 24, 20)(3, 14, 7, 8, 19, 18, 22)(11, 15) \\
15A & 15 & $1^1\cdot 3^1 \cdot 5^1 \cdot 15^1$ & (2, 13, 23, 21, 16)(3, 7, 9, 5, 4, 18, 15, 12, 19, 22, 20, 10, 14, 17, 24)(6, 8, 11) \\
15B & 15 & $1^1\cdot 3^1 \cdot 5^1 \cdot 15^1$ & (2, 23, 16, 13, 21)(3, 12, 24, 15, 17, 18, 14, 4, 10, 5, 20, 9, 22, 7, 19)(6, 8, 11) \\
 & & & \\
\hline
\end{tabular}
}}
\end{center}
\vspace{-0.5cm}
\caption{Conjugacy classes of $M_{23}\subset M_{24}$ (Type 1)} \label{t1}
\renewcommand{\arraystretch}{0.5}
\end{table}

\begin{table}[H]
\renewcommand{\arraystretch}{0.5}
\begin{center}
\vspace{0.5cm}
 {\scriptsize{
\begin{tabular}{|c|c|c|c|}
\hline
 & & & \\
Conjugacy Class & Order & Cycle shape & Cycle \\
\hline
 & & & \\
2B & 4 & $2^{12}$ & (1, 8)(2, 10)(3, 20)(4, 22)(5, 17)(6, 11)(7, 15)(9, 13)(12, 14)(16, 18)(19, 23)(21, 24) \\
3B & 9 & $3^8$ & (1, 10, 3)(2, 24, 18)(4, 13, 22)(5, 19, 15)(6, 7, 23)(8, 21, 12)(9, 16, 17)(11, 20, 14) \\
 & & & \\
\hline
 & & & \\
12B & 144 & $12^2$ & (1, 12, 24, 23, 10, 8, 18, 6, 3, 21, 2, 7)(4, 9, 11, 15, 13, 16, 20, 5, 22, 17, 14, 19) \\
6B &  36 & $6^4$ & (1, 24, 10, 18, 3, 2)(4, 11, 13, 20, 22, 14)(5, 17, 19, 9, 15, 16)(6, 21, 7, 12, 23, 8) \\
4C & 16 & $4^6$ & (1, 23, 18, 21)(2, 12, 10, 6)(3, 7, 24, 8)(4, 15, 20, 17)(5, 14, 9, 13)(11, 16, 22, 19) \\
10A & 20 & $2^2 \cdot 10^2$ & (1, 8)(2, 18, 21, 19, 13, 10, 16, 24, 23, 9)(3, 4, 5, 12, 15, 20, 22, 17, 14, 7)(6, 11) \\
21A & 63 & $3^1 \cdot 21^1$ & (1, 3, 9, 15, 5, 12, 2, 13, 20, 23, 17, 4, 14, 10, 21, 22, 19, 6, 7, 11, 16)(8, 18, 24) \\
21B & 63 & $3^1 \cdot 21^1$ & (1, 12, 17, 22, 16, 5, 23, 21, 11, 15, 20, 10, 7, 9, 13, 14, 6, 3, 2, 4, 19)(8, 24, 18) \\
4A & 8 & $2^4 \cdot 4^4$ & (1, 4, 8, 15)(2, 9, 12, 22)(3, 6)(5, 24, 7, 23)(10, 13)(11, 14, 18, 19)(16, 20)(17, 21) \\
12A & 24 &$2^1\cdot 4^1 \cdot 6^1 \cdot 12^1$ & (1, 15, 8, 4)(2, 19, 24, 9, 11, 7, 12, 14, 23, 22, 18, 5)(3, 13, 20, 6, 10, 16)(17, 21)\\
 & & & \\
\hline
\end{tabular}
}}
\end{center}
\vspace{-0.5cm}
\caption{Conjugacy classes of $M_{24}\not\in M_{23}$ (Type 2)} \label{t2}
\renewcommand{\arraystretch}{0.5}
\end{table}

\section{Twisted Elliptic Genus} \label{twistgenus}

In this section we construct the twisted elliptic genus of the conjugacy classes in table \ref{t1} and 
then for the classes of $2B$ and $3B$ from table \ref{t2}. 
Among the classes in table \ref{t1}, the complete elliptic genus for 
 the classes  $pA$ with
$p = 2, 3, 5, 7$ were given in \cite{David:2006ji}. 
To quote the result we first define
\begin{eqnarray}
 A &=& \left [ \frac{\theta_2(\tau, z)^2  }{\theta_2(\tau, 0)^2}
 + \frac{\theta_3(\tau, z)^2  }{\theta_3(\tau, 0)^2}
 + \frac{\theta_4(\tau, z)^2  }{\theta_4(\tau, 0)^2} \right],  \\  \nonumber
 B(\tau, z) &=& \frac{\theta_1(\tau, z)^2}{\eta(\tau)^6}
\end{eqnarray}
and
\begin{equation}\label{defen}
 {\cal E}_N(\tau) = \frac{12i}{\pi (N-1) } \partial_\tau [ \ln \eta(\tau) - \ln \eta(N\tau) ].
\end{equation}
Under $SL(2, \mathbb{Z})$ transformation $A(\tau, z)$ 
transforms as a weak Jacobi form of weight 0 and index 1 and $B(\tau, z)$ 
transforms  as a weak Jacobi form of weight  $-2$ and index 1. 
Now ${\cal E}_N(\tau)$ transforms as a modular form of
weight $2$ under the group $\Gamma_0(N)$.
Its transformations under $T$ and $S$ transformations of 
$SL(2, \mathbb{Z})$ are given by 
\begin{equation}\label{entrans}
{\cal E}_N(\tau + 1) = {\cal E}_N(\tau) , \qquad  {\cal E}_N (-1/\tau) =  - \frac{\tau^2}{N}
{\cal E}_N(\tau/N) .
\end{equation}
Then the twisted elliptic genera in all the sectors  for the classes $pA$ with 
$p = 2, 3, 5, 7$ are given by 
\begin{eqnarray}\label{pAtwist}
F^{(0,0)}( \tau, z) & =& \frac{8}{N} A(\tau, z), \\ \nonumber
F^{(0, s)} (\tau, z) &=& \frac{8}{N(N+1) } A(\tau, z)  - \frac{2}{N+1} B(\tau, z) {\cal E}_N(\tau) 
, \quad \hbox{for} \, 1\leq s\leq (N-1),  \\ \nonumber
F^{(r, rk)}  &=& \frac{8}{N(N+1) } A(\tau, z)  + 
\frac{2}{N(N+1)} {\cal E}_N\left( \frac{\tau +k}{N} \right) B(\tau, z),  \\ \nonumber
& &  \hbox{for}\, 1\leq r\leq (N-1), 1\leq k\leq (N-1). 
\end{eqnarray}
Note that $rk$ is defined up to mod $N$. Here $N = 2, 3, 5, 7$ corresponding to 
the classes $pA$ respectively.  Let us discuss the low lying coefficients in the expansion 
of $F^{(0, s)}$ which is given by 
\begin{equation}
 F^{(0,s)} ( \tau , z) = \sum_{j \in {\mathbb{Z}}, n =0 }^\infty 
 c^{(0, s) } ( 4n - j^2) e^{2\pi i n \tau} e^{2\pi i j z}.
\end{equation}
Then it is easy to see from (\ref{pAtwist}) that the low lying coefficients  satisfy the following 
property
\begin{equation} \label{c1}
 \sum_{s =0}^{N-1} c^{(0, s)}( \pm 1)  = 2 .
\end{equation}
The above set of equations   corresponds to the number of $(0,0), (0, 2), (2, 0), (2,2)$ forms
of the $pA$ orbifold of $K3$. As expected, these are the same as $K3$, since  the orbifold
preserves these forms \cite{David:2006ud}.  The $(2,0)$ and $(0,2)$ forms are 
holomorphic forms which are required  to be preserved if Type II theory compactified 
on  the $pA$ orbifold $(K3\times T^2)/\mathbb{Z}_N$ needs to be a ${\cal N}=4$ theory. 
The orbifold preserves
the $0$-form as well as the top-form of $K3$. 
In fact the twisted elliptic genus satisfies the  stronger property
 \begin{equation}
 c^{(0, s)}( \pm 1) = \frac{2}{N}, \qquad s = 0, \cdots N-1 .
\end{equation}
Now we can also see that
\begin{equation}
 \sum_{s =0}^{N-1)} c^{(0, 0)} (0)  =  2 \left(
 \frac{24}{N+1} - 2 \right) .
 \end{equation}
The last equation  corresponds to the number of the 
$(1, 1)$ forms which are reduced from the $K3$ value of $20$ to 
$12, 8, 4, 2$  for $N= 2, 3, 5, 7$ respectively.  Finally the orbifold action for all these classes on 
$K3$ produces another $K3$. Therefore, the elliptic genus of $K^3/\mathbb{Z}_N$ should 
be the same as that of $K3$. This implies that we should obtain
\begin{equation}
 \sum_{r, s =0}^{N-1} F^{(r, s)} (\tau, z) = 8 A(\tau, z) .
\end{equation}
Substituting the expressions for the twisted elliptic genus given in (\ref{pAtwist}), 
we see that this is ensured by the following identity
satisfied by ${\cal E}_N(\tau)$ for $N$ prime. 
\begin{equation} \label{enprimen}
\sum_{s=0}^{N-1}{\cal E}_N(\frac{\tau+s}{N})-N {\cal E}_N(\tau)=0 .
\end{equation}

We will first focus on the classes with prime orders, $11A$ and $23A$
and obtain the twisted elliptic genera for these cases. We then move to the
classes with composite order $4A, 6A,  8A, 14A, 14B, 15A, 15B$.
All cases with composite orders involve sub-orbits in the sectors of the twisted elliptic genus. 
To determine the twisted elliptic genus in these sub-orbits we use its  correspondence
with the cycle structure in $M_{24}$. 
Finally we discuss the cases of $2B, 3B$  which are 
automorphisms of order $4, 6$ respectively. For these cases we use quantum symmetry
to determine the twisted elliptic genus in the sub-orbits.

\subsection{The conjugacy class $11A$ and $23A$}
\label{sec11a}
\subsubsection*{11A class}

The twining character for this class was determined in 
\cite{Cheng:2010pq,Eguchi:2010fg,Gaberdiel:2010ch}, 
it is given by  \footnote{We have multiplied   the twining character in  
\cite{Cheng:2010pq,Eguchi:2010fg,Gaberdiel:2010ch} by  
$1/11$. The reason for this normalization will be explained subsequently. }
\begin{eqnarray}
F^{(0, 0)} &=& \frac{8}{11} A(\tau, z),  \\ \nonumber  
 F^{(0,1)} &=& \frac{2}{33}A(\tau, z) -B(\tau, z) 
\left(\frac{1}{6}{\cal E}_{11}(\tau) -\frac{2}{5}\eta^2(\tau)\eta^2(11\tau)\right) .
\end{eqnarray}
To determine the twisted elliptic genus in all the sectors we will use the transformation 
law given in  (\ref{fulltwist}). Since  $A(\tau, z), B(\tau, z)$ are weak Jacobi forms
under $SL(2, \mathbb{Z})$ it is the transformation property of  the 
$\Gamma_0(11)$ forms ${\cal E}_{11}(\tau)$
and $\eta^2(\tau)\eta^2(11\tau)$ in (\ref{entrans})
under $SL(2, \mathbb{Z})$ which allows us to move to the other sectors. 

It is first useful  to show that 
$F^{(0,1)} = F^{(0, s)}$ for all $s = 2, \cdots N -1$. It is important to point out 
if this fact is assumed for prime $N$ then the construction 
of all the sectors  proceeds very  straight forwardly using modular transformations. 
However as we will see we can prove the equality $F^{(0,1)} = F^{(0, s)}$, this 
will involve several steps.
For this we will need the following identities obeyed by the Dedekind $\eta$-function and 
${\cal E}_N$ for $N$ odd. 
\begin{eqnarray} \label{halfshift1}
 \eta(\tau+\frac{1}{2})&=&e^{\pi i/24}\frac{\eta^3(2\tau)}{\eta(\tau)\eta(4\tau)},  \label{etaiden}
 \\ 
 {\cal E}_{N}(\tau+\frac{1}{2})&=&-{\cal E}_{N}(\tau)-4{\cal E}_{N}(4\tau)+6{\cal E}_{N}(2\tau) .
 \label{enindent}
\end{eqnarray}
Note that the first equation is a simple consequence of the definition of the 
$\eta$ function in its product form. The second equation can be obtained
using the first equation and the definition (\ref{defen}).

First let us show    $F^{(0,1)} = F^{(0,2)}$.  Since we will have to keep
track of how the factor ${\cal E}_{11}(\tau)$ and $\eta^2(\tau)\eta^2(11\tau)$ transforms 
under $SL(2, \mathbb{Z})$, let us label these  coefficients in the twisted elliptic genus
in terms of the sector it occurs.
Let 
\begin{equation}
 f^{(0,1)} = {\cal E}_{11} (\tau), \qquad g^{(0, 1)} = \eta^2(\tau)\eta^2(11\tau) .
\end{equation}
Now  under an $S$ transformation we know that from (\ref{fulltwist}) we obtain the 
$F^{(1,0)}$ sector or the $F^{(10, 0)}$ sector. Therefore  using the transformation 
in (\ref{entrans}), we obtain 
\begin{equation}
 F^{(1, 0)}(\tau, z) = F^{(10, 0)} = \frac{2}{33} A(\tau, z) + B(\tau, z) \left( 
 \frac{1}{66} {\cal E}_{11} (\frac{\tau}{11} ) - \frac{2}{55} \eta^2(\frac{\tau}{11} )\eta^2(\tau)\right) .
\end{equation}
We label the coefficients
\begin{equation}
 f^{(1,0)}(\tau)  = f^{(10, 0)} =\frac{1}{\tau^2} {\cal E}_{11}( \frac{\tau}{11} ) , \qquad g^{(1,0)}(\tau) = g^{(10, 0)} =
\frac{1}{\tau^2} \eta^2(\frac{\tau}{11} )\eta^2(\tau) .
\end{equation}
Now from the  $F^{(1, 0)}$ sector we can obtain the $F^{(1, s)}$ sector by a $(T)^s$ transformation 
and it results in 
\begin{equation}
 F^{(1, s)}(\tau, z) = \frac{2}{33} A(\tau, z) + B(\tau, z) \left( 
 \frac{1}{66} {\cal E}_{11} (\frac{\tau + s}{11} ) - 
 \frac{2}{55} \eta^2(\frac{\tau +s}{11} )\eta^2(\tau+s)\right) .
\end{equation}
An equation identical to the above exists for the sectors $F^{(10, s)}$. 
Thus the coefficients 
\begin{equation}
 f^{(1,s)}(\tau )   = f^{(10, 10s)}(\tau) = {\cal E}_{11}( \frac{\tau+s}{11} ) , 
 \qquad g^{(1,s)}(\tau)  = g^{(10, 10s)}(\tau)=
 \eta^2(\frac{\tau+s}{11} )\eta^2(\tau+s )
\end{equation}

Let us now move to the $F^{(2, s)}$ sectors. For this we begin with 
$F^{(1,2)}$ and perform the $S$ transformation. 
From (\ref{fulltwist}) we see that we can obtain $F^{(2, 10)}$ sector. 
The $S$ transformation  also relates $F^{(1,2)} = F^{(9 , 1)}$ and the conclusions
we obtain for the $F^{(2, 10)}$  sector also holds for the $F^{(9 , 1)}$ sector. 
This sector will have the coefficients of $B$ given by 
\begin{eqnarray}
 f^{(2, 10)}(\tau) &=& 
 \frac{1}{\tau^2} {\cal E}_{11} ( - \frac{1}{11\tau} + \frac{2}{11} ) , \\  \nonumber
 g^{(2, 10)}(\tau) &=& \frac{1}{\tau^2}
 \eta^2( -\frac{1}{11\tau}  +\frac{2}{11} )\eta^2(-\frac{1}{\tau} +2) .
\end{eqnarray}
However, this is not useful, since we would like to obtain a $q$ expansion for these twisted sectors. 
We will  establish the identity
\begin{eqnarray} \label{man0}
 f^{(2, 10)} (\tau) = \frac{1}{\tau^2} {\cal E}_{11} ( \frac{\tau + 5}{11} )  = f^{(1, 5)} (\tau), \\ \nonumber
g^{(2, 10)} (\tau) = \frac{1}{\tau^2}  \eta^2(\frac{\tau+5}{11} )\eta^2(\frac{\tau+5}{11} ) 
= g^{(1, 5)} (\tau) .
\end{eqnarray}
This enables us to perform the $q$ expansion in these sectors. 
To begin, we see that using the transformations under $S$ and $T$  (\ref{entrans}) 
we obtain 
\begin{equation} \label{man1}
 {\cal E}_{11} ( - \frac{1}{11\tau} + \frac{2}{11} ) = - \frac{11}{( 2 - 1/\tau)^2}
 {\cal E}_{11} ( \frac{1}{2} - \frac{1}{ 2( 2\tau -1) } ) .
\end{equation}
We can use the the identity (\ref{enindent}) to remove the shift by $1/2$. This leads to 
\begin{equation}\label{man2} 
 {\cal E}_{11} ( - \frac{1}{11\tau} + \frac{2}{11} ) = 
 - {\cal E}_{11} ( - \frac{1}{2 ( 2\tau - 1) } ) - 4 {\cal E}( - \frac{2}{ ( 2\tau - 1) } ) 
 + 6 {\cal E}_{11} ( - \frac{1}{ 2 \tau - 1} ) .
\end{equation}
Using the $S$ transformation on the LHS of (\ref{man2})    leads us to 
\begin{equation}
  {\cal E}_{11} ( - \frac{1}{11\tau} + \frac{2}{11} )  = \frac{( 2\tau -1)^2 }{11}
  \left[ 4 {\cal E}_{11} ( 2 \frac{ ( 2 \tau -1)}{11} )  + {\cal E}_{11}(  \frac{ 2 \tau -1}{22 } ) 
   - 6 {\cal E}_{11} ( \frac{  2\tau - 1}{11} )  \right] .
\end{equation}
Substituting  the above equation into (\ref{man1}) we obtain 
\begin{eqnarray} \label{man3} 
  {\cal E}_{11} ( - \frac{1}{11\tau} + \frac{2}{11} )  &=& \tau^2 
  \left[ - 4 {\cal E}_{11} \left( 4 \frac{ ( 2 \tau -1)}{22} \right )  -  {\cal E}_{11}\left(  \frac{ 2 \tau -1}{22 } \right) 
   + 6 {\cal E}_{11} ( \frac{  2\tau - 1}{11} )  \right],  \nonumber \\
   &=&\tau^2 {\cal E}_{11} \left( \frac{\tau  + 5}{11} \right) .
\end{eqnarray}
In the last line we have again used the identity (\ref{enindent}) with $\tau$ replaced by $ (2\tau - 1)/22$. 
Therefore using (\ref{man3}) we see that the first identity in equation (\ref{man0}) is true.
Exactly a similar manipulation involving $g^{(2, 10)}$ but using (\ref{etaiden}) to remove
shifts by $1/2$ will enable us to prove the second equation in (\ref{man0}). 
Thus we obtain the relation 
\begin{equation}\label{man5}
 F^{(2, 10)} = F^{(1, 5)} 
\end{equation}
Now using the $T$ transformation and the fact that $N=11$ is prime we can obtain all the twisted sectors $F^{(2, s)}$ as well as $F^{(9, s)}$. 

To show that $F^{(0, 1)} = F^{(0, 2)}$, we start with (\ref{man5}) and perform 
a $(T)^6$ transformation to both sides of the equation using (\ref{fulltwist}). 
Then we obtain  $F^{(2, 0)} = F^{(1, 0)}$ and thus by a $S$ transformation we see that
$F^{(0, 1)} = F^{(0, 2)}$. 

Similar manipulations allow us to obtain all the other sectors. 
Let us briefly go over one more case. We start with $F^{(2, 4)}$ which can be obtained by 
performing a $T^{-5}$ transformation. Thus the 
coefficient of the $B$ term is given by 
\begin{equation}
 f^{(2, 4)}(\tau)  =\frac{1}{\tau^2} {\cal E}_{11} ( \frac{\tau + 2}{11} ) 
\end{equation}
Lets do the $S$ transformation on $F^{(2, 4)}$ with will take us to 
either $F^{(4, 9)}$ or $F^{(7, 2)}$ and the coefficient of the $B$ term in this sector 
contains 
\begin{equation}
f^{(4, 9)} (\tau) = \frac{1}{\tau^2} {\cal E}_{11} (  - \frac{1}{11\tau} + \frac{2}{11}  ) .
\end{equation}

Therefore using the same set of manipulations we see that 
$F^{(4, 9)} = F^{(7, 2)} = F^{(1, 5)}$. 
Therefore by a first $T^{-(5)}$ and then an $S$ action we can show 
$F^{(0, 4) } = F^{( 0, 7)} = F^{(0,1)}$. 
Using these steps we obtain the relations among different sectors as given in the following chart.
 \begin{table}[H]
 \renewcommand{\arraystretch}{0.5}
 \begin{center}
 \vspace{0.5cm}
  {\small{
 \begin{tabular}{|c|c|c|c|}
 \hline
  & S & $T^{(-5)}$ & S  \\
 \hline
 &  & & \\
$ F^{(1,2)} $ & $ F^{(2, 10)} = F^{(9, 1)}$ {\bf=} $ F^{(1, 5)} $ & $ F^{(2, 0)} = F^{(9, 0)}= F^{(1, 0)}= F^{(10, 0)}$ & $ F^{(0,2)} = F^{(0, 9)}= F^{(0, 1)}= F^{(0, 10)}$ \\
& & & \\
$ F^{(2,4)} $ & $ F^{(4, 9)} = F^{(7, 2)}$ {\bf=} $ F^{(1, 5)} $ & $ F^{(4, 0)} = F^{(7, 0)}= F^{(2, 0)}$ & $ F^{(0,4)} = F^{(0, 7)}= F^{(0, 2)}$ \\
& & & \\
$ F^{(4,8)} $ & $ F^{(8, 7)} = F^{(3, 4)}$ {\bf=} $ F^{(1, 5)} $ & $ F^{(8, 0)} = F^{(3, 0)}= F^{(4, 0)}$ & $ F^{(0,8)} = F^{(0, 3)}= F^{(0, 4)}$ \\
& & & \\
$ F^{(3,6)} $ & $ F^{(6, 8)} = F^{(5, 3 )}$ {\bf=} $ F^{(1, 5)} $ & $ F^{(6, 0)} = F^{(5, 0)}= F^{(3, 0)}$ & $ F^{(0,6)} = F^{(0, 5)}= F^{(0, 3)}$ \\
 \hline
 \end{tabular}
 }}
 \end{center}
 \vspace{-0.5cm}
 \caption{Table showing the S and T transformation starting from the twisted sector given in the leftmost coloumn.}
 \renewcommand{\arraystretch}{0.5}
 \end{table}
Going through these steps we obtain the following formula for the 
twisted elliptic genus for $11A$. 
\begin{eqnarray}\label{answ11A}
F^{(0, 0)} &=& = \frac{8}{11} A(\tau, z), \\ \nonumber 
F^{(0,s)}&=&\frac{2}{33}A(\tau, z)-B (\tau, z) 
\left(\frac{1}{6}{\cal E}_{11}(\tau) -\frac{2}{5}\eta^2(\tau)\eta^2(11\tau)\right) ,\\ \nn
F^{(r, rs ) }&=&\frac{2}{33}A(\tau, z) +B(\tau, z) 
\left(\frac{1}{66}{\cal E}_{11}(\frac{\tau+s}{11})-\frac{2}{55}
\eta^2(\tau+s)\eta^2(\frac{\tau+s}{11})\right) .
\end{eqnarray}
 The low lying values of the twisted elliptic genus for the $11A$ case satisfies
  \begin{eqnarray}\label{c11a}
 && c^{(0, s)}( \pm 1) = \frac{2}{11}, \qquad s = 0, \cdots N-1,  \\ \nonumber
 && \sum_{s=0}^{N-1}  c^{(0, s)}( \pm 1)  = 2 .
\end{eqnarray}
Thus type II compactifications on such the orbifold $(K3\times T^2)/\mathbb{Z}_N$
where $\mathbb{Z}_N$ acts as the $11A$ automorphism on $K3$ together 
with a $1/11$ shift on one of the circles of $T^2$  preserves ${\cal N}=4$ supersymmetry.   
Note that the normalization of multiplying the twining character given in 
\cite{Cheng:2010pq,Eguchi:2010fg,Gaberdiel:2010ch}
by $1/11$ is to ensure the condition given in (\ref{c11a}). 
However the number of $(1,1)$ forms preserved by the orbifold vanishes as 
\begin{equation}
\sum_{s=0}^{N-1}  c^{(0, s)}( 0) =0.
\end{equation}
Thus even the K\"{a}hler form of $K3$ is projected out, which implies this 
orbifold is not geometric. 
We can evaluate the elliptic genus of the $11A$ orbifold of $K3$, the result is given by 
\begin{equation}
\sum_{r, s=0}^{N-1} F^{(r, s)} (\tau, z) = 8 A(\tau, z).
\end{equation}
We have verified this equation by performing a  $q$-expansion of both the 
left hand side as well as the right hand side. 
The fact that the elliptic genus of the $11A$ orbifold of $K3$ satisfies this identity 
implies that the $11A$ orbifold of $K3$ is $K3$ itself.

The  structure of all the twisted sectors is similar to that seen in the $pA$ cases 
in (\ref{pAtwist}).  In fact  for the cases in $pA$,  if one is  given just the twining elliptic genus 
we can obtain the complete twisted elliptic genus using the same manipulations 
discussed for the $11A$.

\subsubsection*{23A/B class}
The twining characters for the conjugacy classes 23A and 23B are identical and was 
determined in \cite{Cheng:2010pq,Eguchi:2010fg,Gaberdiel:2010ch}. It is given by 
\begin{eqnarray} \label{23ab}
F^{(0, 0)} &=& \frac{8}{23} A(\tau, z),  \\ \nonumber  
 F^{(0,1)} &=& \frac{1}{69}A-B \left(\frac{1}{12}{\cal E}_{23}-\frac{1}{22}f_{23,1}(\tau)
 -\frac{7}{22}\eta^2(\tau)\eta^2(23\tau) \right).
\end{eqnarray}
We can use the same procedure as discussed for the class $11A$ in the previous section 
to determine the twisted elliptic genus in all the sectors. 
Essentially we  use the transformation 
law given in  (\ref{fulltwist})  to move to twisted elliptic genus in the other sectors from the 
$(0,1)$ sector. As discussed in the  previous section we need identities 
satisfied by  the modular forms 
${\cal E}_{23}(\tau)$,  $ \eta^2(\tau)\eta^2(23\tau) $ and $f_{23,1}(\tau)$ to express the expansion 
in terms of $e^{-2\pi i /\tau}$  in terms of a the usual $q$ expansion. 
Note that all these transform as  modular forms under $\Gamma_0(23)$. 
The identities analogous to equation (\ref{man3}) can be found by similar manipulations. 
The new form  $f_{23,1}(\tau)$
under $\Gamma_0(23)$ has been constructed in \cite{zagier,link} which involves 
Hecke eigenforms. 
A closed formula for $f_{23,1}(\tau)$ in terms of $\eta$ functions 
 is provided in the ancillary files associated 
with \cite{Gaberdiel:2012gf}. This is given by  
\begin{equation}\label{f321}
f_{23,1}(\tau)=2\frac{\eta^3(\tau)\eta^3(23\tau)}{\eta(2\tau)\eta(46\tau)}+8\eta(\tau)\eta(2\tau)\eta(23\tau)\eta(46\tau)+8 \eta^2(2\tau)\eta^2(46\tau)+5\eta^2(\tau)\eta^2(23\tau)
\end{equation}
%
It can be seen that from   (\ref{f321}) that the  $S$ transformation of $f_{23,1}(\tau)$ is given by 
\be
f_{23,1}(-\frac{1}{\tau})=-\frac{\tau^2}{23}f_{23,1}(\frac{\tau}{23}).
\ee
To obtain the $q$ expansion of the various sectors of the twisted elliptic genus 
we need the following identities to be satisfied by $f_{23,1}(\tau)$. 
{\footnotesize{ \begin{eqnarray} \label{f23iden1}
  f_{23,1}(-\frac{1}{23\tau}+\frac{1}{23}) &=&{\tau^2}f_{23,1}(\frac{\tau+22}{23}), \label{red} \\
   f_{23,1}(-\frac{1}{23\tau}+\frac{2}{23})&=&{\tau^2}f_{23,1}(\frac{\tau+11}{23}), \label{blue} \\
   f_{23,1}(-\frac{1}{23\tau}+\frac{3}{23}) &=& {\tau^2}f_{23,1}(\frac{\tau+15}{23}) \label{green}
\end{eqnarray}} }
{ \footnotesize{\begin{eqnarray} \label{f23iden2}
f_{23,1}(-\frac{1}{23\tau}+\frac{4}{23})={\tau^2}f_{23,1}(\frac{\tau+17}{23}), \quad 
f_{23,1}(-\frac{1}{23\tau}+\frac{5}{23})={\tau^2}f_{23,1}(\frac{\tau+9}{23})\\ \nn
f_{23,1}(-\frac{1}{23\tau}+\frac{6}{23})={\tau^2}f_{23,1}(\frac{\tau+19}{23}), \quad 
f_{23,1}(-\frac{1}{23\tau}+\frac{7}{23})={\tau^2}f_{23,1}(\frac{\tau+13}{23})\\ \nn
f_{23,1}(-\frac{1}{23\tau}+\frac{8}{23})={\tau^2}f_{23,1}(\frac{\tau+20}{23}), \quad 
f_{23,1}(-\frac{1}{23\tau}+\frac{10}{23})={\tau^2}f_{23,1}(\frac{\tau+16}{23})\\ \nn
f_{23,1}(-\frac{1}{23\tau}+\frac{12}{23})={\tau^2}f_{23,1}(\frac{\tau+21}{23}), \quad
f_{23,1}(-\frac{1}{23\tau}+\frac{14}{23})={\tau^2}f_{23,1}(\frac{\tau+18}{23}).
              \end{eqnarray} } }
              A similar analysis given in the previous section 
              can be used to prove these identities. 
              We have not done this, but have numerically verified these identities. 
Again following  a similar analysis to   ${\cal E}_{11}(\tau)$ given in   the previous section  
we have  shown 
that 
the  modular form ${\cal E}_{23}(\tau)$ as well 
as $\eta^2(\tau)\eta^2(23\tau)$ obey the identities  (\ref{f23iden1}) and (\ref{f23iden2}). 
Therefore combining the modular transformation obeyed by the twisted elliptic genus
(\ref{fulltwist}) as well as the identities in (\ref{f23iden1}) and (\ref{f23iden2}) 
 we obtain  the twisted elliptic genus of the conjugacy class
$23A$. The result is given by 
\bea \label{23abtwist}
F^{(0,k)}(\tau, z) &=&\frac{1}{23}\left(\frac{1}{3}A-B\left(\frac{23}{12}{\cal E}_{23}(\tau) -\frac{23}{22}f_{23,1}(\tau)-\frac{161}{22}\eta^2(\tau)\eta^2(23\tau)\right)\right) ,\\ \nn
F^{(r, rk) }(\tau, z) &=&\frac{1}{23}
\left[\frac{1}{3}A+B\left(\frac{1}{12}{\cal E}_{23}(\frac{\tau+k}{23})-
\frac{1}{22}f_{23,1}(\frac{\tau+k}{23})-\frac{7}{22}\eta^2(\tau+k)\eta^2(\frac{\tau+k}{23})\right)\right].
\eea
The low lying coefficients of this twisted elliptic genus satisfy
\begin{eqnarray}
 && c^{(0, s)}( \pm 1) = \frac{2}{23}, \qquad s = 0, \cdots N-1,  \\ \nonumber
 && \sum_{s=0}^{N-1}  c^{(0, s)}( \pm 1)  = 2
\end{eqnarray}
As we have discussed earlier, this implies that type II compactifications on 
the orbifold $K3\times T^2/\mathbb{Z}_N$ preserves ${\cal N}=4$ supersymmetry. 
We also have 
\begin{equation}
 \sum_{s=0}^{N-1} c^{(0, s)}(0) = -2.
\end{equation}
When the RHS side of this equation is a positive integer,  it corresponds to 
the number of $(1,1)$ forms preserved by the orbifold. Here,  we obtain a 
result which is a negative integer, the orbifold is therefore  not geometric. 
Just as in the case of the $11A$ orbifold,  the  elliptic genus  of $23A/23B$ orbifold 
reduces to that of $K3$. This can be seen by showing  the twisted 
elliptic genera of the $23A/23B$ orbifold satisfies 
\be
\sum_{r,s=0}^{N-1} F^{(r,s)}(\tau, z) =8A(\tau,z).
\ee
We have verified this identity by substituting the twisted elliptic genus from (\ref{23abtwist}) and 
performing the $q$ expansion on both sides of the 
above equation.

\subsection{Automorphisms $g'$  with composite order and $g'\in M_{23}$}

Let us consider automorphisms $g'$ with composite order and those which belong 
to $M_{23}\subset M_{24}$.  Examples of these are the classes $4B, 6A, 8A, 14A, 15A$
given in table \ref{t1}. 
When the order of the automorphism $g'$ is composite, 
we cannot use  the $SL(2, \mathbb{Z})$ 
  modular transformation in (\ref{fulltwist}) to arrive at all the sectors 
 of the twisted elliptic genus started from the twining character. 
 For example for the case of $4B$ which is of the order $4$ we cannot 
 reach the sectors $(0, 2), (2, 0),  (2, 2)$ starting from the twining character $(0, 1)$. 
 We call these sectors sub-orbits. 
 In general if the order $N$ admits a factorization 
 \begin{equation}
  N = \prod_{i} n_i 
 \end{equation}
then there is a sub-orbit for each divisor. 
Since the sub-orbits are not accessible by modular transformations from the 
twining character $(0,1)$ one needs to make a choice of a particular character in these sectors. 
To be more specific, 
consider the sub-orbit corresponding to the divisor $n_i$ we need to make a choice for the character
\begin{eqnarray}
 F^{(0, n_i)} (\tau_i) =\frac{1}{N}{\rm Tr}_{R\, R }[(-1)^{F_{K3}+\bar{F}_{K3} }g'^{n_i} e^{2\pi i zF_{K3}}
 q^{L_0-c/24} \bar{q}^{\bar{L_0}-\bar{c}/24}].
\end{eqnarray}
We will see in all the cases for composite orders with $g'\in M_{23} \subset M_{24}$,  we will see that
from the cycle shape of  $g'^{n_i}$ corresponds to a conjugacy class of 
order $N/n^i$. 
Therefore by appealing to Mathieu moonshine symmetry 
we can choose for $F^{(0, n_i)} (\tau, z)$, the  twining character corresponding to the
conjugacy class with the cycle structure of $g'^{n_i}$. We  show that with these choices
we can complete the construction of the twisted elliptic genera for the remaining 
conjugacy classes in table \ref{t1}.

\subsection*{$4B$ class}

The twining character for the $4B$ conjugacy class is given by 
\begin{equation}
 F^{(0, 1)}(\tau, z)  = \frac{A(\tau, z)}{3} -  \frac{B(\tau, z) }{4} \left( - \frac{1}{2} {\cal E}_2(\tau) 
 + 2 {\cal E}_4(\tau) \right) . 
\end{equation}
Since the modular forms involved in the twining character is in $\Gamma_{0}(N)$, the order of the 
automorphism corresponding to the $4B$ class is $4$. 
Therefore the sectors $(0,2), (2, 0), (2, 2)$ are not accessible using $SL(2, \mathbb{Z})$ modular
transformations.   Now the cycle shape of $g'$ in this class is given by 
$1^4\cdot 2^2 \cdot 4^4$ and the cycle shape of $g'^2$ is given by $1^8 \cdot 2^8$.
From table \ref{t1} we see that 
this cycle shape coincides with the conjugacy class $2A$. 
Therefore we choose for the twisted elliptic genus in the sector $(0, 2)$ to be identical to be the $1/2$ the 
twisting character $(0, 1)$  of the $2A$ conjugacy class.  
The choice of normalization is because we are in an order $4$ conjugacy class. 
We will also show that this normalization results in the expected values for the low lying coefficients
of the elliptic genus. 
Similarly sectors $(2, 0)$ and the $(2, 2)$   of the $4B$ conjugacy class coincide with
$1/2$ the twisted sectors $(1,0)$ and $(1, 1)$ of the $2A$ class. 
The rest of the sectors can be determined by using the relation (\ref{fulltwist}) and identities 
relating expansions in $e^{-2\pi i /\tau}$ to $e^{2\pi i \tau}$. 
For this we need the following identities 
\bea\label{e2e4iden}
{\cal E}_2(\tau+1/2)&=&-{\cal E}_2(\tau)+2{\cal E}_2(2\tau) ,\\ \nn
{\cal E}_4(\tau+1/2)&=&\frac{1}{3}(-{\cal E}_2(\tau)+4{\cal E}_2(2\tau)) .
\eea
One can  prove these identities using the the definition of ${\cal E }_N(\tau)$ in 
(\ref{defen}) together with the first equation of (\ref{halfshift1}).  
The  identities in (\ref{e2e4iden}) 
allow us to obtain the $(2, 1)$ or the $(2, 3)$ sector from the $(1,2)$ using 
a similar analysis followed in section \ref{sec11a}. 
The result for the twisted elliptic genus using these inputs is given by 
\bea \label{twch4b}
F^{(0,0)}(\tau, z) &=&2A(\tau, z) ,\nn \\ 
 F^{(0,1)}(\tau, z) &=& F^{(0,3)}(\tau, z) =
 \frac{1}{4}\left[\frac{4A}{3}-B \left(-\frac{1}{3}{\cal E}_2(\tau)+2{\cal E}_4(\tau) \right)\right], \\ \nn
 F^{(1,s)}(\tau, z) &=& F^{(3,3s)}=\frac{1}{4}\left[\frac{4A}{3}+
 B \left(-\frac{1}{6}{\cal E}_2(\frac{\tau+s}{2})+\frac{1}{2}{\cal E}_4(\frac{\tau+s}{4}) \right)\right], \\ \nn
 F^{(2,1)}(\tau, z) &=& F^{(2,3)}=\frac{1}{4}\left(\frac{4A}{3}
 -\frac{B}{3} (3{\cal E}_2(\tau)-4{\cal E}_2(2\tau)\right) ,\\ \nn
 F^{(0,2)}(\tau, z) &=&\frac{1}{4}\left(\frac{8A}{3}-\frac{4B}{3}{\cal E}_2(\tau)\right) ,\\ \nn
 F^{(2,2s)}(\tau, z) &=&\frac{1}{4}\left(\frac{8A}{3}+\frac{2B}{3}{\cal E}_2(\frac{\tau+s}{2})\right) .
\eea
Note that sector $(0,1)$  is the 
twining character given by \cite{Cheng:2010pq,Eguchi:2010fg,Gaberdiel:2010ch} 
for the $4B$ conjugacy  class. Using this,  the modular transformation property 
(\ref{fulltwist})  and the relations in (\ref{e2e4iden})  we obtain the sectors 
$(2, 1), (2, 3)$. Finally the sectors $(0, 2), (2, 2s)$ belong to the sub-orbit  which can be identified 
with the $2A$ class. Note the twisted elliptic genus for this sub-orbit is $1/2$ of that
twisted elliptic genus for the $2A$ class. 
It is interesting to note that   our result in (\ref{twch4b}) for the twisted elliptic genus 
coincides with that obtained in  \cite{Govindarajan:2009qt}. This 
was obtained prior to the discovery of the $M_{24}$ symmetry. 
The approach followed in \cite{Govindarajan:2009qt} involved writing down the 
possible $\Gamma_0(4)$  and  $\Gamma_0(2)$ forms allowed in the 
$(0, s)$ sectors and constraining  the coefficients using topological data. 

Let us now evaluate the low lying coefficients of the elliptic genus
We have 
\begin{eqnarray}
 && c^{(0, s)}( \pm 1) = \frac{1}{2}, \qquad s = 0, \cdots N-1,  \\ \nonumber
 && \sum_{s=0}^{N-1}  c^{(0, s)}( \pm 1)  = 2
\end{eqnarray}
and 
\begin{equation}
 \sum_{s=0}^{N-1} c^{(0, s)}(0) =  6 .
\end{equation}
This equation implies that the number of $(1, 1)$ forms due to the  orbifolding is down to $6$ from 
$20$ of the $K3$. This agrees with the analysis of  \cite{Chaudhuri:1995dj} which studies the orbifold of 
$K3$ dual to the $N=4$ CHL compactification.  We can therefore identify the compactification of
type II on 
$(K3\times T^2)/\mathbb{Z}_4$ where $\mathbb{Z}_4$ is the $4A$ automorphism to be dual to 
the $N=4$ heterotic CHL compactification. 
Let us now evaluate the full elliptic genus of $K3$ orbifolded  by the $4A$ automorphism. 
This is given by 
\begin{equation}
 \sum_{r, s=0}^{N-1} F^{(r, s)}(\tau, z) = 8 A(\tau, z) .
\end{equation}
To show this we substitute the twisted elliptic genus given in (\ref{twch4b}) along with  the 
identity  in (\ref{enprimen}) and finally use the relation
\begin{equation}
 \frac{1}{4}\sum_{s=0}^3{\cal E}_4(\frac{\tau+s}{4}) = {\cal E}_2 (\tau) .
\end{equation}

\subsection*{6A class}

The twining character for th $6A$ conjugacy class is given by 
\begin{eqnarray}\label{tw6a}
 F^{(0,1)} &=&\frac{1}{6}\left(\frac{2A}{3}-B \left(-\frac{1}{6}{\cal E}_2(\tau)
-\frac{1}{2}{\cal E}_3(\tau)+\frac{5}{2}{\cal E}_6(\tau) \right)\right)
\end{eqnarray}
From this, it is easy to see that that $6A$ automorphism is of the order $6$, which 
admits $2$ and $3$ as non-trivial divisors. 
Therefore there are 2 independent sub-orbits of orders $3$ and $2$ respectively.
These sub-orbits 
cannot be accessed using $SL(2, \mathbb{Z}) $ modular transformations from the $(0,1)$ sector. 
The  sub-orbits are  the following twisted sectors
\begin{eqnarray}
 & &a: \,    ( 0,2), (0, 4); ( 2, 0), (2, 2), (2, 4); (4, 0), ( 4,2), (4, 4), \\ \nonumber
 & &b: \,   (0, 3); (3, 0), (3, 3).
\end{eqnarray}
Now to determine the twisted elliptic genus in the sub-orbit $a$, first examine the 
$(0, 2)$ sector.  The cycle shape of $g'$ for the $6A$ conjugacy class can be 
read out from the table \ref{t1} and is given by $1^2 \cdot 2^2 \cdot 3^2 \cdot 6^2$. 
The  cycle shape of  $g^{\prime 2}$  for $6A$
 is given by  $1^6 \cdot 3^6$ which is identical to that cycle 
shape of the conjugacy class $3A$. Therefore we take the twisted elliptic genus of 
for the sub-orbit $(a)$ to be   $1/2$ of  the twisted elliptic genera of 
the $3A$ class. Similarly for the sub-orbit $(b)$ the cycle shape is obtained by looking
at $g^{\prime 3}$ which is $1^8 \cdot 2^8$. This coincides with 
the cycle structure of the $2A$ conjugacy class. Therefore for the twisted 
elliptic genera  of the sub-orbit $(b)$ we can take 
$1/3$ the twisted elliptic genera of the $2A$ conjugacy class. 

The sectors other  than the sub-orbits $(a)$ and $(b)$ can be 
reached using the $SL(2, \mathbb{Z})$ transformation given in (\ref{fulltwist}). 
Again to convert expansions in $e^{ -2\pi i /\tau}$ to 
expansions in $e^{2\pi i \tau}$ we need the following identity
obtained by using (\ref{defen}) and the first equation of (\ref{halfshift1}).
\be\label{e6half}
{\cal E} _6(\tau+\frac{1}{2})=\frac{1}{5}\left(-{\cal E}_2(\tau)+2{\cal E}_2(2\tau)+4{\cal E}_3(2\tau)\right).
\ee
Lets illustrate this in obtaining the $(3, 1)$ or the $(3, 5)$ sectors from the $(0, 1)$ sector. 
First using the $S$ transformation on the twining character in (\ref{tw6a}) 
we obtain the $(1, 0)$ sector which is given by 
\begin{eqnarray}
F^{(1, 0)}(\tau, z)&=&F^{(5, 0)}=\frac{1}{6}\left[\frac{2A}{3}+B \left(-\frac{1}{12}{\cal E}_2(\frac{\tau}{2})-\frac{1}{6}{\cal E}_3(\frac{\tau}{3})+\frac{5}{12}{\cal E}_6(\frac{\tau}{6}) \right)\right].
\nonumber \\
\end{eqnarray}
Then using $T^3$ transformation we can reach the $(1, 3)$ sector, which is given by 
\begin{eqnarray}\label{man6b1}
F^{(1, 3)}&=&F^{(5, 3)}=\frac{1}{6}\left[\frac{2A}{3}+B \left(-\frac{1}{12}{\cal E}_2(\frac{\tau+3}{2})-\frac{1}{6}{\cal E}_3(\frac{\tau+3}{3})+\frac{5}{12}{\cal E}_6(\frac{\tau+3}{6}) \right)\right]. \nonumber
\\
\end{eqnarray}
We can now use the $S$ transformation, to obtain the $(3, 1)$ or the $(3, 5)$ 
sectors.  It is easy to see from the argument of ${\cal E}_6$ in (\ref{man6b1}) 
we will require the identity in (\ref{e6half}) to perform the $S$ transformation. 
The relations we need are
\bea
{\cal E} _6(\frac{\tau}{6}+\frac{1}{2})&=&\frac{1}{5}\left(-{\cal E}_2(\tau/6)+2{\cal E}_2(\tau/3)+4{\cal E}_3(\tau/3)\right), \\ \nn
{\cal E} _6(\frac{-1}{6\tau}+\frac{1}{2})&=&\frac{1}{5}\left(-{\cal E}_2(-1/6\tau)+2{\cal E}_2(-1/3\tau)+4{\cal E}_3(-1/3\tau)\right) ,\\ \nn
&=& \frac{\tau^2}{5}(18{\cal E}_2(3\tau)-12{\cal E}_3(\tau)-9{\cal E}_2(\frac{3\tau}{2})).
\eea
This results in the following expression for the $(3, 1)$ or the $(3, 5)$ sector
\begin{equation}
F^{(3, 1)} (\tau, z) = F^{(3, 5)}(\tau, z) = \frac{A }{9}-\frac{B}{12}{\cal E}_3(\tau)-\frac{B}{72}{\cal E}_2(\frac{\tau+1}{2})+\frac{B}{8}{\cal E}_2(\frac{3\tau+1}{2}).
\end{equation}
From the $(3, 1)$ sector by performing the  $T$ and then the $S$ transformation
we can reach the $(2, 3)$ sector and again we will require the use of the 
identity (\ref{e6half}) as well as 
\begin{equation}
{\cal E}_2( -\frac{1}{2\tau} + \frac{1}{2} ) = \tau^2 {\cal E}_2 ( \frac{\tau + 1}{2} ) .
\end{equation}
Finally from the $(2, 3)$ sector by $T$ transformations we can reach the 
$(2, 1)$ sector as well as the $(2, 5)$ sector.

Using all these inputs  the sectors of the twisted elliptic 
genus for $6A$ are given by 
\bea\label{6a1}
F^{(0,0)}&=&\frac{4}{3}A;\quad  F^{(0,1)}=  F^{(0,5)};\quad  F^{(0,2)}=  F^{(0,4)};\\ \nn
F^{(0,1)} &=&\frac{1}{6}\left[\frac{2A}{3}-B \left(-\frac{1}{6}{\cal E}_2(\tau)
-\frac{1}{2}{\cal E}_3(\tau)+\frac{5}{2}{\cal E}_6(\tau) \right)\right],\\ \nn
 F^{(0,2)}&=& \frac{1}{6}\left[2A-\frac{3}{2}B {\cal E}_3(\tau)\right], \\ \nn
 F^{(0,3)}&=& \frac{1}{6}\left[\frac{8A}{3}-\frac{4}{3}B {\cal E}_2(\tau)\right].
\eea
\bea \label{6a2} \nn
F^{(1, k)}&=&F^{(5, 5k)}=\frac{1}{6}\left[\frac{2A}{3}+B \left(-\frac{1}{12}{\cal E}_2(\frac{\tau+k}{2})-\frac{1}{6}{\cal E}_3(\frac{\tau+k}{3})+\frac{5}{12}{\cal E}_6(\frac{\tau+k}{6}) \right)\right], \\
\eea
\bea\label{6a3}
F^{(2, 2k+1)}&=& \frac{A}{9}+\frac{B}{36}{\cal E}_3(\frac{\tau+2+k}{3}), \\ \nn
F^{(4, 4k+1)}&=& \frac{A}{9}+\frac{B}{36}{\cal E}_3(\frac{\tau+1+k}{3}), \\ \nn
F^{(3, 1)}&=&F^{(3, 5)}= \frac{A}{9}-\frac{B}{12}{\cal E}_3(\tau)-\frac{B}{72}{\cal E}_2(\frac{\tau+1}{2})+\frac{B}{8}{\cal E}_2(\frac{3\tau+1}{2}), \\ \nn
F^{(3, 2)}&=&F^{(3, 4)}= \frac{A}{9}-\frac{B}{12}{\cal E}_3(\tau)-\frac{B}{72}{\cal E}_2(\frac{\tau}{2})+\frac{B}{8}{\cal E}_2(\frac{3\tau}{2}), \\ \nn
\eea
\bea\label{6a4}
 F^{(2r,2rk)}&=& \frac{1}{6}\left[2A+\frac{1}{2}B {\cal E}_3(\frac{\tau+k}{3})\right],  \nn \\
  F^{(3,3k)}&=& \frac{1}{6}\left[\frac{8A}{3}+\frac{2}{3}B {\cal E}_2(\frac{\tau+k}{2})\right].
\eea

 The low lying coefficients of the $6A$ twisted elliptic genus
is given by 
\begin{eqnarray}
 && c^{(0, s)}( \pm 1) = \frac{1}{3}, \qquad s = 0, \cdots 5,  \\ \nonumber
 && \sum_{s=0}^{5}  c^{(0, s)}( \pm 1)  = 2 .
\end{eqnarray}
and 
\begin{equation}
 \sum_{s=0}^{5} c^{(0, s)}(0) =  4 .
\end{equation}
Therefore the  number of $(1, 1)$ forms  is $4$. This agrees 
 with   \cite{Chaudhuri:1995dj} which studies the orbifold of 
$K3$ dual to the $N=6$ CHL compactification.  We  therefore identify the compactification of
type II on 
$(K3\times T^2)/\mathbb{Z}_6$ where $\mathbb{Z}_6$ is the $6A$ automorphism to be dual to 
the $N=4$ heterotic CHL compactification. 
The  full elliptic genus of $K3$ orbifolded  by the $6A$ automorphism. 
is given by 
\begin{equation}
 \sum_{r, s=0}^{N-1} F^{(r, s)}(\tau, z) = 8 A(\tau, z) .
\end{equation} 
Thus the result of the $6A$ orbifold of $K3$ is $K3$ itself.

\subsection*{8A class}

The twining character for the $8A$   conjugacy  class is given by 
\begin{equation}
F^{(0,1)}= 
\frac{1}{8}\left[\frac{2A}{3}-B \left(-\frac{1}{2}{\cal E}_4(\tau)+\frac{7}{3}{\cal E}_8(\tau) \right)\right].
\end{equation}
Therefore the  $8A$ automorphism in $K3$ is of the order $8$, this admits $2$ and 
$4$ as non-trivial divisors. The independent sub-orbits are of the 
length $4$ and $2$ respectively. 
From table \ref{t1}, the cycle shape of the conjugacy class for $8A$ is given by 
$1^2 \cdot 2^1 \cdot 4^1 \cdot 8^2$. 
The cycle shape of $g^{\prime 2}$ is $1^4 \cdot 2^2 \cdot 4^4$ which is 
identical to the $4B$ conjugacy class. 
The sectors of the twisted elliptic genus belonging to this sub-orbit 
of order $4$ 
\begin{eqnarray}
& & (0, 2), (0, 4), (0, 6); \\ \nonumber
 && (2, 0), (2, 2), (2, 4), (2, 6); \\ \nonumber
 && (4, 0), (4, 2), (4, 4), (4, 6); \\ \nonumber
 && ( 6, 0), (6, 2), (6, 4), (6, 6).
\end{eqnarray}
Since the $4A$ already has a sub-orbit of order  $2$ which coincides with the 
sub-orbit of of the $8A$ conjugacy class we do not need to consider this 
sub-orbit independently. 
Thus the twisted sectors in this sub-orbit is taken to be $1/2$ that of the 
$4B$ conjugacy class given in (\ref{twch4b}). 
The rest of the sectors can be determined by using the modular transformation
(\ref{fulltwist}) and the identity 
\begin{equation}
{\cal E}_8(\tau+1/2)=\frac{1}{7} \left(-{\cal E}_2(\tau)+2{\cal E}_2(2\tau)+6{\cal E}_4(2\tau)\right).
\end{equation}
Going through a similar analysis as in the case of $6A$ we obtain 

\bea
F^{(0, 0)} (\tau, z) &=& A(\tau, z), \\ \nonumber
 F^{(0,1)}&=& F^{(0,3)}= F^{(0,5)}= F^{(0,7)},\\ \nn
&=&\frac{1}{8}
\left[\frac{2A}{3}-B \left(-\frac{1}{2}{\cal E}_4(\tau)+\frac{7}{3}{\cal E}_8(\tau) \right)\right].
\eea

\be
F^{(r, rk)}(\tau, z) =\frac{1}{8}\left[\frac{2A}{3}+\frac{B}{8} \left(-{\cal E}_4(\frac{\tau+k}{4})+\frac{7}{3}{\cal E}_8(\frac{\tau+k}{8})\right)\right].
\ee
where $r=1,3,5,7$.

\bea
 F^{(2,1)}&=& F^{(6,3)}= F^{(2,5)}= F^{(6,7)},\\ \nn
&=& \frac{1}{8}\left[\frac{2A}{3}+\frac{B}{3} \left(-{\cal E}_2(2\tau)+\frac{3}{2}{\cal E}_4(\frac{2\tau+1}{4}) \right)\right];\\ \nn
 F^{(2,3)}&=& F^{(6,5)}= F^{(2,7)}= F^{(6,1)},\\ \nn
&=& \frac{1}{8}\left[\frac{2A}{3}+\frac{B}{3} \left(-{\cal E}_2(2\tau)+\frac{3}{2}{\cal E}_4(\frac{2\tau+3}{4}) \right)\right].
\eea
\bea
 F^{(0,2)}&=& F^{(0,6)}=\frac{1}{8}\left(\frac{4A}{3}-B \left(-\frac{1}{3}{\cal E}_2(\tau)+2{\cal E}_4(\tau) \right)\right),\\ \nn
 F^{(0,4)}&=&\frac{1}{8}\left(\frac{8A}{3}-\frac{4B}{3}{\cal E}_2(\tau)\right),\\ \nn
 F^{(2,2s)}&=& F^{(6,6s)}=\frac{1}{8}\left(\frac{4A}{3}+B \left(-\frac{1}{6}{\cal E}_2(\frac{\tau+s}{2})+\frac{1}{2}{\cal E}_4(\frac{\tau+s}{4}) \right)\right),\\ \nn
 F^{(4,4s)}&=&\frac{1}{8}\left(\frac{8A}{3}+\frac{2B}{3}{\cal E}_2(\frac{\tau+s}{2})\right),\\ \nn
 F^{(4,2)}&=& F^{(4,6)}=\frac{1}{8}\left(\frac{4A}{3}-\frac{B}{3} (3{\cal E}_2(\tau)-4{\cal E}_2(2\tau)\right),\\ \nn
 F^{(4,2k+1)}&=&\frac{1}{8}\left(\frac{2A}{3}+B \left(\frac{4}{3}{\cal E}_2(4\tau)-\frac{2}{3}{\cal E}_2(2\tau)-\frac{1}{2}{\cal E}_4(\tau) \right)\right).
\eea

Finally the low lying coefficients of this orbifold 
satisfy
\begin{eqnarray}
 && c^{(0, s)}( \pm 1) = \frac{1}{4}, \qquad s = 0, \cdots 7,  \\ \nonumber
 && \sum_{s=0}^{7}  c^{(0, s)}( \pm 1)  = 2.
\end{eqnarray}
and 
\begin{equation}
 \sum_{s=0}^{7} c^{(0, s)}(0) =  2 .
\end{equation}
The above equation implies that the number of $(1, 1)$ forms is $2$ which 
agrees with the $K3$ orbifold dual to the $N=8$ CHL compactification
 \cite{Chaudhuri:1995dj} . 
The  full elliptic genus of $K3$ orbifolded  by the $8A$ automorphism. 
is given by 
\begin{equation}
 \sum_{r, s=0}^{N-1} F^{(r, s)}(\tau, z) = 8 A(\tau, z) 
\end{equation} 
Thus the elliptic genus of the  $8A$ orbifold of $K3$ is $K3$ itself.

We have seen that the $K3$ orbifold by the $4A$, $6A$ and the $8A$ conjugacy class
can be identified with type II on  $(K3\times T^2)/\mathbb{Z}_N$ orbifolds dual to $N=4, 6, 8$
CHL compactifications   of the heterotic string respectively. 
Therefore  with the result for 
the twisted elliptic genus for these cases along with the twisted
elliptic genus for the $pA$ cases with $p= 2, 3, 5, 7$ completes the 
analysis of the twisted elliptic genus for all the $7$ CHL compactifications
discussed in \cite{Chaudhuri:1995dj}. 

There are $2$ remaining conjugacy classes in table (\ref{t1}). 
These are the $14A$ and the $15A$ conjugacy classes. The
 construction
for the twisted elliptic genus for these classes proceeds along similar lines
as that discussed for the composite orders. The result is 
given in appendix \ref{frslist}. 

\subsection{Automorphisms $g'$  with composite order and $g'\not \in M_{23}$ }

The conjugacy  classes listed in table \ref{t2} are all of composite orders. 
Therefore they admit sub-orbits under the action of $SL(2, \mathbb{Z})$. 
However the cycle shape  in the sub-orbit is not unique enough  to determine
the twisted elliptic genus. 
For instance consider the conjugacy class $2B$ in table \ref{t2}, squaring $g'$ leads to a 
the cycle shape of the identity class. 
Recently \cite{Gaberdiel:2013psa}  constructed  an explicit rational conformal field theory consisting 
of $6$ $SU(2)$ WZW models at level 1 which realizes $K3$. The
action of the orbifold by  $g'$ belonging to the conjugacy class $2B$ is explicitly realized 
in this CFT. 
It was observed that  this orbifold satisfied  the property called 
`quantum symmetry'
\begin{equation} \label{qs}
\sum_{r, s=0}^{N-1} F^{(r, s)}( \tau, z ) = 0 
\end{equation}
In this section starting from  the twining characters
for the $2B$ and $3B$ conjugacy class given in \cite{Gaberdiel:2013psa} 
we determine all the sectors of the twisted elliptic genus. 
This is done  by assuming quantum symmetry together 
with the  following condition on the low lying coefficients
of the twisted elliptic genus
\begin{equation}\label{susy}
\sum_{s =0}^{N-1} c^{(0, s)} (\pm 1) =  2 
\end{equation}
where order $N$ is $4, 9$ for the $2B$ and $3B$ conjugacy class respectively. 
As we have discussed earlier, 
the above condition on the low lying coefficients of the twisted elliptic genus 
ensures that the type II theory compactified on $(K_3\times T^2)/\mathbb{Z}_N$
preserves ${\cal N}=4$ supersymmetry.

\subsection*{Twisted elliptic genus of 2B}

An explicit realization of the $2B$ orbifold of $K3$ was given  in \cite{Gaberdiel:2013psa}
in which $K3$ is realized a rational CFT consisting of $6$ $SU(2)$ WZW models
at level $1$. 
Rather than use this realization, we will start from 
the twining character given in \cite{Cheng:2010pq,Eguchi:2010fg,Gaberdiel:2010ch} 
for the $2B$ conjugacy class
\begin{equation}
F^{(0,1)}(\tau, z)  =\frac{B(\tau, z) }{2}({\cal E}_2(\tau)-{\cal E}_4(\tau))
\end{equation}
Note that this is distinct from the classes belonging to table \ref{t1} in that it does 
not have any component of the weak Jacobi form $A(\tau, z)$. 
It is clear form the structure of the twining character,  the $2B$  automorphism 
is or the order $4$. 
Using the modular transformations (\ref{fulltwist}) together with the 
identities in (\ref{e2e4iden}),   we can determine the elliptic genus in 
the following sectors to be given by 
\begin{eqnarray}
F^{(0,1)}(\tau, z) &=&  F^{(0,3)}(\tau, z), \\ \nonumber
F^{(1,s)}(\tau, z) &=& F^{(3,3s)}(\tau, z) =-\frac{B(\tau, z)}{4}({\cal E}_2(\frac{\tau+s}{2})-{\cal E}_4(\frac{\tau+s}{4}))
, \\ \nonumber
F^{(2,1)}(\tau, z) &=&F^{(2,3)}(\tau, z) =\frac{B(\tau, z)}{2}(-\frac{1}{6}{\cal E}_2(\tau)+\frac{2}{3}{\cal E}_2(2\tau))
\end{eqnarray}
The remaining sectors $(0, 2), (2, 0), (2, 2)$ belong to a sub-orbit. 
To determine the structure of the elliptic genus in this sub-orbit let us first 
focus on the $(0,2)$ sector. 
We assume that is 
 a $\Gamma_0(2)$ weak Jacobi form. Thus it can be written as 
\begin{equation}
F^{(0,2)}(\tau, z) =\alpha A(\tau, z) +\beta B(\tau, z)  {\cal E}_2(\tau).
\end{equation}
where $\alpha, \beta$ are undetermined constants. 
Now the sectors   $(2, 0)$  and  $(2, 2)$ can be determined using the 
modular transformations  (\ref{fulltwist}) to be 
\begin{equation}
F^{(2, 2s)} (\tau, z) = \alpha A(\tau, z)  - \frac{\beta}{2} {\cal E}_2 ( \frac{\tau + s}{2} ) .
\end{equation}
Imposing the equations (\ref{qs}) and (\ref{susy}) 
we obtain 
\begin{equation}
\alpha = \beta =  - \frac{2}{3}.
\end{equation}

To summarize the twisted elliptic genus for the $2B$ conjugacy class is given by
\bea\label{comp2b}
F^{(0,0)}(\tau, z) &=&2A;\quad  F^{(0,1)}(\tau, z) =  F^{(0,3)}(\tau, z), \\ \nn
F^{(0,1)}(\tau, z) &=&\frac{B(\tau, z) }{2}({\cal E}_2(\tau)-{\cal E}_4(\tau)),\\ \nn
F^{(0,2)}(\tau, z)  &=&-\frac{2A(\tau, z) }{3}-\frac{2B(\tau, z) }{3}{\cal E}_2(\tau) ,\\ \nn
F^{(1,s)}(\tau, z) &=& F^{(3,3s)}=-\frac{B(\tau, z) }{4}({\cal E}_2(\frac{\tau+s}{2})-
{\cal E}_4(\frac{\tau+s}{4})),\\ \nn
F^{(2,1)}(\tau, z)  &=&F^{(2,3)}=\frac{B(\tau, z) }{2}(-\frac{1}{6}{\cal E}_2(\tau)+
\frac{2}{3}{\cal E}_2(2\tau)),\\ \nn
F^{(2,2s)}(\tau, z)  &=&-\frac{2A(\tau, z) }{3}+\frac{B(\tau, z) }{3}{\cal E}_2(\frac{\tau+s}{2}) . \\ \nn
\eea
We have also  evaluated the complete twisted  elliptic genus using the 
explicit rational CFT realization of this 
orbifold in \cite{Gaberdiel:2013psa} and have verified that it agrees with that 
given in (\ref{comp2b}). 
Evaluating 
the low lying coefficient corresponding to the invariant $(1, 1)$ forms 
of $K3$ we obtain 
\begin{equation}
\sum_{s =0}^{3} c^{(0, s)} ( 0) = 0.
\end{equation}
Therefore, as expected  the orbifold corresponding to the $2B$ conjugacy class
is non-geometric.

\subsection*{Twisted elliptic genus of 3B}

Among the conjugacy classes in table \ref{t2}) the $3B$ 
class  can also  be completely determined using the quantum symmetry (\ref{qs}) 
and the supersymmetry condition (\ref{susy}). 
The twining character in this class is given by 
\begin{equation}\label{3btwin}
 F^{(0,1)}(\tau, z) =-\frac{2B(\tau, z) }{9}\frac{\eta^6(\tau)}{\eta^2(3\tau)}
\end{equation}
From the modular properties of the $\eta$ function it is clear that the 
$3B$ automorphism is of order $9$. 
The following sectors
\begin{eqnarray} \label{suborbit3b}
 && (0, 3), (0, 6); \\ \nonumber
 & &(3, 0), (3, 3), (3, 6); \\ \nonumber
 && (6, 0), (6, 3), (6, 6) 
\end{eqnarray}
forms a sub-orbit under $Sl(2, \mathbb{Z})$ modular transformations. 
The remaining sectors can be obtained  from the 
twining character in (\ref{3btwin}) by using the transformation (\ref{fulltwist}),
together with the modular properties of the $\eta$ function. 
Once that is obtained we assume the following Jacobi  weak form of $\Gamma_0(3)$
for the  $(0, 3)$ sector of the (\ref{suborbit3b}). 
\begin{equation}
F^{(0,3)}(\tau, z) =\alpha A(\tau, z) +\beta B (\tau, z) {\cal E}_3(\tau).
\end{equation}
Here $\alpha, \beta$ are undetermined constants. 
Then using modular transformation (\ref{fulltwist})  and the identities  in (\ref{identenn}) for $N=3$ 
we can obtain the twisted elliptic genus in the sub-orbit. 
Finally imposing the conditions (\ref{qs}) and (\ref{susy}) we determine the constants
$\alpha, \beta$ as
\begin{equation}
 \alpha = -\frac{1}{9}, \qquad \beta = \frac{1}{4}.
\end{equation}

Using all these steps we obtain the twisted elliptic genus  for the $3B$ conjugacy class to be given by 
\bea
F^{(0,0)} (\tau, z) &=&\frac{8A(\tau, z) }{9},\\ \nonumber 
F^{(0,1)} (\tau, z) &=&  F^{(0,2)}= F^{(0,4)}= F^{(0,5)}= F^{(0,7)}= F^{(0,8)};\\ \nn
F^{(0,1)}(\tau, z)  &=&-\frac{2B(\tau, z) }{9}\frac{\eta^6(\tau)}{\eta^2(3\tau)},\\ \nn
F^{(0,3)}(\tau, z)  &=&-\frac{A(\tau, z) }{9}-\frac{B(\tau, z) }{4}{\cal E}_3(\tau) ,\\ \nn
F^{(r,rs)}(\tau, z)  &=&\frac{2B(\tau, z) }{3}\frac{\eta^6(\tau+s)}{\eta^2(\frac{\tau+s}{3})}, \quad\quad r=1,2,4,5,7,8 \\ \nn
F^{(3,1)}(\tau, z)  &=&-\frac{2B(\tau, z)}{9}e^{2\pi i/3}\frac{\eta^6(\tau)}{\eta^2(3\tau)} ,\\ \nn
&=&  F^{(3,4)}=F^{(3,7)}= F^{(6,2)}=F^{(6,8)}=F^{(6,5)};\\ \nn
F^{(3,2)}(\tau, z)  &=&-\frac{2B(\tau, z) }{9}e^{4\pi i/3}\frac{\eta^6(\tau)}{\eta^2(3\tau)},
\\ \nn
&=& F^{(3,5)}=F^{(3,8)} = F^{(6,1)}=F^{(6,7)}=F^{(6,4)}; \\ \nn
F^{(3r,3rk)}(\tau, z)  &=&-\frac{A(\tau, z) }{9}+\frac{B(\tau, z) }{12}{\cal E}_3(\frac{\tau+k}{3}) .\\ \nn
\eea
The number of $(1, 1)$ forms is given by 
consider the following low lying coefficients of the twisted elliptic genus 
\begin{equation}
 \sum_{s =0}^{8} c^{(0, s)} ( 0) = -2
\end{equation}
Again the orbifold of $K3$ by the $3B$ conjugacy class is non-geometric.

The rest of the conjugacy classes in table (\ref{t2}) have more than one sub-orbits. 
quantum symmetry given in (\ref{qs}) and the supersymmetry condition (\ref{susy}) 
is not enough to determine the unknown constants in these sub-orbits. It will be 
interesting to determine the twisted elliptic genera for all the remaining conjugacy 
classes of table \ref{t2}.

\subsection{Comparision with literature} \label{comparision}

As remarked earlier the work of \cite{Gaberdiel:2012gf} provides the  mathematical justification 
for the construction of the twisted elliptic genus over for all the cyclic orbifolds considered 
in this paper. 
To compare with the results of this paper, let us briefly review their construction. 
Let $g'$ be the cyclic orbifold corresponding to the conjugacy 
class of $M_{24}$ with order $N$.  Then the twisted elliptic genus admits the following 
decomposition in terms of the characters of the ${\cal N}=4$ 
superconformal algebra with central charge $c=6$
\begin{equation}\label{chrexp}
 F^{(r, s)} (\tau, z) = \sum_{k = n + \frac{r}{N}\geq 0 }^\infty {\rm Tr}_{{\cal H}_{g^{\prime r}, k}   }  
 ( \rho_{g^{\prime r} , k} (g^{\prime s}) ) {\rm ch}_{ h = \frac{1}{4} +k, l  } (\tau, z) 
\end{equation}
Note that $l = \frac{1}{2}$ except when $h =1/4$ for which both $l = \frac{1}{2} $ and $l =0$ 
are understood to be present in the sum. 
The vector space ${\cal H}_{g^{\prime r}, k}$ is finite dimensional and is the 
projective representation of the centralizer $C_{M_{24}}(g')$ which satisfies 
 properties detailed in \cite{Gaberdiel:2012gf}. 
 Thus the problem of determining the twisted twining elliptic genera 
 reduces to determining characters of the projective representations. 
 Though  not easy to extract from the   ancillary files   provided along   with \cite{Gaberdiel:2012gf}, 
 a careful examination 
 of the files lists out  
 some of  the    twisted twining 
 elliptic genera for  the orbifolds considered in the paper.  Notably it is  only the $F^{(1,s)}$ sector 
 which is listed out  in the ancillary files. 
 The files also enable the evaluation of the characters of the projective 
 representations and a verification of the expansion of the twisted 
 elliptic genera as given in (\ref{chrexp}). 
 However explicit expressions such as that given in 
 equations (\ref{pAtwist}), (\ref{answ11A}), (\ref{6a1}) - (\ref{6a4})
 are not listed in the main body of the paper 
 \footnote{ We have been informed by Mathias Gaberdiel that these explicit
 expressions were known to the authors of \cite{Gaberdiel:2012gf} however they did not write them out in 
 the body of  their paper.}.  
 To arrive at these 
 reasonably compact expressions we had to perform modular transformations and use 
  identities such as (\ref{man3}), (\ref{f23iden1}), (\ref{e6half}). These identities were demonstrated for all the 
  cases except for the orbifold by the class $23A/B$.  Let us also emphasize that 
  the $F^{(1, s)}$ sector which are provided 
  in the ancillary files associated with \cite{Gaberdiel:2012gf} are the simplest to obtain by 
  modular transformations. 
  We have also not made any assumption that the $F^{(0,1)}$ sector is the 
  same as the $F^{(0, s)}$ sector for $s\neq 1$ and $N$ prime, but  have arrived at it as a consequence
  of the identities derived in this paper as can be seen from the detailed discussion of the 
  case of $11A$ class. 
 
 Note that explict formulae for the 
 twisted elliptic genera for  $pA$ orbifolds with $p =2, 3, 5, 7$  were
 known even before the discovery of moonshine symmetry in \cite{David:2006ji} and before the work of 
 \cite{Gaberdiel:2012gf}. 
 Since the latter  paper  as well the present work uses modular 
 transformations (\ref{fulltwist}) to obtain the twisted elliptic genera we are assured that 
 both the constructions agree.  
 In our discussion we have also explicitly compared the low lying coefficients of the 
 twisted elliptic genera for the case of $4B, 6A, 8A$ orbifolds with the Hodge numbers
 of the CHL compactifications discussed in \cite{Chaudhuri:1995fk,Chaudhuri:1995dj} and found agreement. 
 The check that sum over all the sectors of the orbifolds when $g' \in M_{23}$ yields 
 back the elliptic genus of $K3$ was also performed in \cite{Gaberdiel:2012gf} 
 as can be read from the discussion in the  text  \footnote{See discussion in the beginning of 
 section 4. of \cite{Gaberdiel:2012gf}.}. 
 In this work this is assured by the identities of the kind 
 given (\ref{enprimen}). 

As we have discussed earlier, when  the order of the orbifold is composite,  there are 
sub-orbits in the twisted sectors which cannot be reached by modular transformations
from the twining character. 
We have used moonshine symmetry to determine the twisted elliptic genus in these sectors. 
The treatment of such  situations in \cite{Gaberdiel:2012gf} is more general. Their discussion also encompasses 
orbifolds by non-cyclic groups.  Though not treated explicitly, the 
case of the cyclic orbifolds is implicit in their discussion \footnote{We thank 
Mathias Gaberdiel for correspondence which enabled us to compare our work with 
\cite{Gaberdiel:2012gf}.}.

\section{$1/4$ BPS dyon partition functions} \label{dydegen}

Given the twisted elliptic genus one can construct a Siegel modular form as follows
\cite{David:2006ud}. 
The twisted elliptic genus can be expanded as
\begin{equation}
F^{(r,s)}(\tau, z) = \sum_{b=0}^1 \sum_{j \in 2\mathbb + b , n \in \mathbb{Z}/N}
c_{b}^{(r, s)} ( 4n - j^2)  e^{ 2\pi i n \tau + 2\pi i j z} .
\end{equation}
Then a Siegel modular form associated with the twisted elliptic genus is given by 
\begin{eqnarray}\label{siegform}
\tilde{\Phi}(\rho,\sigma,v)&=&e^{2\pi i(\tilde{\al}\rho+\tilde{\beta}\sigma+v)}\\ \nn
&&\prod_{b=0,1}\prod_{r=0}^{N-1}
\prod_{\begin{smallmatrix}k'\in \mathcal{Z}+
\frac{r}{N},l\in \mathcal{Z},\\ j\in 2\mathcal{Z}+b\\ k',l\geq0, j<0 k'=l=0\end{smallmatrix}}
(1-e^{2\pi i(k'\sigma+l\rho+jv)})^{\sum_{s=0}^{N-1}e^{2\pi isl/N}c_b^{r,s}(4k'l-j^2)}.
\end{eqnarray}
where 
\bea
\tilde{\beta}&=&\frac{1}{24N}\chi(M) ,\\ \nn
\tilde{\al}&=&\frac{1}{24N}\chi(M)-\frac{1}{2N}\sum_{s=0}^{N-1}Q_{0,s}\frac{e^{-2\pi is/N}}{1-e^{2\pi is/N}} , \\ \nn
Q_{r,s}&=& N(c_0^{r,s}(0)+2c_1 ^{r,s}(-1)).
\eea
Evaluating $\tilde\alpha, \tilde \beta $ for the twisted elliptic genus corresponding to all the 
conjugacy classes considered in the previous section as well as the $pA$ classes with 
$p = 1, 2, 3, 5, 7$ we obtain 
\begin{equation}
 \tilde \alpha = 1, \qquad \tilde \beta = \frac{1}{N}.
\end{equation}
Here $N$ is the order of the orbifold action. 
This Siegel modular form in (\ref{siegform}) transforms as  a weight $k$ form under appropriate 
sub-groups of $Sp(2, \mathbb{Z})$. The weight $k$ is related to the low lying coefficients
of the twisted elliptic genus and is given by 
\begin{equation}
 k=\frac{1}{2}\sum_{0}^{N-1}c_0 ^{0,s}(-1) .
\end{equation}
The weights of the Siegel modular forms corresponding to the twisted elliptic genera
constructed in this paper is listed in table \ref{t3} and \ref{t4}.  
\begin{table}[H] 
\renewcommand{\arraystretch}{0.5}
\begin{center}
\vspace{-0.5cm}
\begin{tabular}{|c|c|c|c|c|c|c|}
\hline
 & & & & & & \\
Type 1 & pA & 4B & 6A & 8A & 14A & 15A \\
 & & & & & & \\
\hline
 & & & & & & \\
Weight  & $\frac{24}{p+1}-2$ & 3 & 2 & 1 & 0 & 0 \\
\hline
\end{tabular}
\end{center}
\vspace{-0.5cm}
\caption{Weight of Siegel modular forms corresponding to classes in $M_{23}$}
\label{t3}
\renewcommand{\arraystretch}{0.5}
\end{table}
\begin{table}[H]
\renewcommand{\arraystretch}{0.5}
\begin{center}
\vspace{-0.5cm}
\begin{tabular}{|c|c|c|}
\hline
  & & \\
Type 2 &  2B &  3B \\
\hline
 & & \\
Weight  & 0  & -1 \\
 & & \\
\hline
\end{tabular}
\end{center}
\vspace{-0.5cm}
\caption{Weight of Siegel modular forms corresponding to the classes $\not\in M_{23}$}
\label{t4}
\renewcommand{\arraystretch}{0.5}
\end{table}
Now consider type II theory compactified on $(K3\times T^2)/\mathbb{Z}_N$ where 
$\mathbb{Z}_N$ acts as the automorphism $g'$ 
belonging to any of the conjugacy classes together
with a $1/N$ shift along one of the circles of $T^2$, $S^1$. 
Then by the analysis in \cite{David:2006ud},  the generating function 
of the index of $1/4$ BPS states in this theory is given by $1/\tilde{\Phi}(\rho, \sigma, v)$. 
Let us work in the dual heterotic frame in which the orbifolded  heterotic theory is compactified
in general on $T^6$. For example the cases of the $pA$ orbifolds of $K3\times T^2$ with
$p= 2, 3, 4, 5, 6, 7, 8$ corresponds to the $N=2, 3, 4, 5, 6, 7, 8$ CHL compactifications 
on the heterotic side. 
Let us label the charges of the $1/4$ BPS state by $(Q, P)$ corresponding to the 
electric and magnetic charge of the dyon.  Let $Q^2, P^2$ and $Q\cdot P$ denote the 
continuous T-duality invariants in this duality frame. 
Then the $1/4$ BPS index in this frame is given by 
\begin{eqnarray}\label{indexb6}
-B_6(Q, P)  = \frac{1}{N} ( -1)^{ Q\cdot P +1} 
\int_{{\cal C}} d\rho d\sigma d v \; e^{-\pi i ( N \rho Q^2 \sigma P^2/N + 2 v Q\cdot P ) }
\frac{1}{ \tilde \Phi ( \rho, \sigma , v) }.
\end{eqnarray}
The contour ${\cal C}$ is defined over a 3 dimensional subspace of the 
3 complex dimensional space $(\rho = \rho_1 + i \rho_2, \sigma = \sigma_1 + i \sigma_2,
v = v_1 + i v_2 ) $. 
\begin{eqnarray}
\rho_2 = M_1, \qquad \sigma_2 = M_2, \qquad v_2 = - M_3, \\ \nonumber
0\leq \rho_1 \leq 1, \qquad 0 \leq \sigma_1 \leq N, \qquad 0 \leq v_1 \leq 1.
\end{eqnarray}
The choice of $(M_1, M_2, M_3)$ is determined by the domain in which 
one needs to evaluate the index $-B_6$ \cite{Sen:2007vb,Dabholkar:2007vk}. 
We pick up the Fourier coefficients   by expanding 
$1/\tilde \Phi$ in powers of $e^{2\pi i \rho}, e^{2\pi i \sigma}$ and 
$e^{-2 \pi i v}$.  For this expansion to make sense we must have 
\cite{David:2006ud,Sen:2007vb}
\begin{equation} \label{domain}
M_1, M_2 >>0,  \quad M_3 <<0, \qquad |M_3| << M_1, M_2 .
\end{equation}
Since this is an index, the Fourier coefficient $-B_6$ must be an integer. 
Let us now focus on $1/4$ BPS states which are single centered 
black holes. 
Then from the fact that the single centered black holes 
carry zero angular momentum,  it is predicted that the index
$-B_6$ for these black holes is positive \cite{Sen:2010mz}. 
The argument for this goes as follows. 
Given the domain (\ref{domain}), 
these $1/4$ BPS black states have regular event horizons and 
are single centered only if 
  the charges satisfy  the condition \cite{Sen:2010mz}
\begin{equation}\label{pcond}
Q\cdot P \geq 0 , \qquad (Q\cdot P )^2 < Q^2 P^2,  \qquad 
Q^2 , P^2 >0 .
\end{equation}
Thus if we can show that  the index $-B_6$  is positive 
for states satisfying the condition (\ref{pcond}), then 
it will imply the $-B_6$  is positive for single centered 
$1/4$ BPS dyons as predicted from black hole considerations. 
In the next section we show that for low lying charges satisfying (\ref{pcond}),  
$-B_6$ is indeed positive  for all  the Siegel modular forms associated with the 
twisted elliptic genera constructed in this paper. 
This  is the generalization of the observation seen first in 
\cite{Sen:2010mz} for the $pA$ conjugacy classes with $p=1, 2, 3, 5, 7$. 
For $p=1$ and for  a special class of  charges it was proved that the 
coefficient $-B_6$ is positive \cite{Bringmann:2012zr}. 

Before we proceed we will study 2 properties of the Siegel modular forms
which are theta lifts of the twisted elliptic genera constructed in this paper. 
First the Siegel modular forms  factorize in the $v\rightarrow 0$ limit as 
\begin{equation}\label{tildefac}
\lim_{v\rightarrow 0}\tilde \Phi_k(\rho, \sigma, v)  \sim  v^2 f^{(k+2)} (\rho) g^{(k+2)}(\sigma).
\end{equation}
where $f^{(k+2)}, g^{(k+2)}$ are weight $k +2$  modular forms transforming 
under $\Gamma_0(N)$. 
The explicit modular forms on which the $\Phi_k's$ factorize are given in 
table \ref{t5}. 
The function $ 1/f^{(k+2)}(\rho) $ is the partition function 
of purely electric states while $ 1/ g^{(k+2)}(\sigma) $ is the partition 
function of purely magnetic states.  In fact $f^{(k+2)}$ and $g^{(k+2)}$ are related 
to each other by a $S$ transformation. 

\begin{table}[H]
\renewcommand{\arraystretch}{0.5}
\begin{center}
\vspace{0.5cm}
\begin{tabular}{|c|c|c|c|}
\hline
 & & & \\
Conjugacy Class & $k$ & $f^{(k+2)} (\rho)$ & $g^{(k+2)}(\sigma)$\\ \hline
 & & & \\
$p$A & $\frac{24}{p+1}-2$ &$\eta^{k+2}(\rho)\eta^{k+2}(p\rho)$  & $\eta^{k+2}(\sigma)\eta^{k+2}(\sigma/p)$\\
 & & & \\
4B & 3 & $\eta^4(4\rho)\eta^2(2\rho)\eta^{4}(\rho)$ &  $\eta^{4}(\frac{\sigma}{4})\eta^{2}(\frac{\sigma}{2})\eta^{4}(\sigma)$\\
 & & & \\
6A & 2& $\eta^2(\rho)\eta^2(2\rho)\eta^2(3\rho)\eta^2(6\rho)$ & $\eta^2(\sigma)\eta^2(\frac{\sigma}{2}) \eta^2(\frac{\sigma}{3})\eta^2(\frac{\sigma}{6})$\\
 & & & \\
8A & 1 &  $\eta^2(\rho)\eta(2\rho)\eta(4\rho)\eta^2(8\rho)$ & $\eta^2(\sigma)\eta(\frac{\sigma}{2}) \eta(\frac{\sigma}{4})\eta^2(\frac{\sigma}{8})$\\
 & & & \\
14A & 0 &  $\eta(\rho)\eta(2\rho)\eta(7\rho)\eta(14\rho)$ &  $\eta(\sigma)\eta(\frac{\sigma}{2}) \eta(\frac{\sigma}{7})\eta(\frac{\sigma}{14})$\\
 & & &\\
15A & 0 & $\eta(\rho)\eta(3\rho)\eta(5\rho)\eta(15\rho)$ &  $\eta(\sigma)\eta(\frac{\sigma}{3}) \eta(\frac{\sigma}{5})\eta(\frac{\sigma}{15})$\\
 & & & \\
\hline
 & & & \\
2B & 0 & { \Large{ $\frac{\eta^8(\rho)}{\eta^4(2\rho)}$}} &{ \Large{ $\frac{\eta^8(\sigma)}{\eta^4(\frac{\sigma}{2})}$}}\\
 & & & \\
3B & -1 & {\Large{$\frac{\eta^3(\rho)}{\eta(3\rho)}$}} & {\Large{ $\frac{\eta^3(\sigma)}{\eta(\frac{\sigma}{3})}$}}\\
 & & & \\
\hline
\end{tabular}
\end{center}
\vspace{-0.5cm}
\caption{Factorization of $\tilde \Phi_k(\rho, \sigma, v) $ as $\lim v\rightarrow 0$ as shown in (\ref{tildefac}), $p\in \{1,2,3,5,7,11,23\}$} \label{t5}
\renewcommand{\arraystretch}{0.5}
\end{table}

The second property we discuss is the asymptotic property of the 
index in (\ref{indexb6})  when the charges  $Q, P$ are equally large. 
The procedure to obtain the asymptotic behaviour has been 
developed in 
\cite{David:2006yn,David:2006ud,LopesCardoso:2006ugz}, which we summarize  briefly 
\footnote{We follow the discussion in \cite{David:2006ud}.}.
Consider another Siegel modular form $\hat \Phi(\rho, \sigma, v)$
 of weight $k$ associated 
with the twisted elliptic genus defined by 
\bea \label{phihat}
&&\hat{\Phi}(\rho,\sigma,v)=e^{2\pi i(\hat{\al}\rho+\hat{\beta}\sigma+v)}\\ \nn
&&\prod_{b=0,1}\prod_{r=0}^{N-1}\prod_{\begin{smallmatrix}k',l\in \mathcal{Z},\\
 j\in 2\mathcal{Z}+b\\ k',l\geq0, j<0 k'=l=0\end{smallmatrix}}(1-e^{2\pi i r/N}e^{2\pi i(k'\sigma+l\rho+jv)})^{\sum_{s=0}^{N-1}e^{-2\pi isr/N}c_b^{0,s}(4k'l-j^2)}.
\eea
Here we have,
\be
\hat{\beta}=\hat{\al}=\hat{\gamma}=\frac{1}{24}\chi(M)=1
\ee
Under $v\rightarrow 0$, this modular form factorizes  symmetrically in $\rho$ and 
$\sigma$ as 
\begin{equation}\label{hatphifac}
\lim_{v\rightarrow 0} \hat{\Phi}(\rho,\sigma,v) \sim v^2 h^{(k+2)} (\rho) h^{(k+2)}(\sigma)
\end{equation}
Then the leading behaviour of the index $-B_6$ is given by 
\begin{equation}
-B_6 (Q, P) \sim \exp ( -S(Q, P) )
\end{equation}
where $S(Q, P)$ is obtained by minimizing the function 
\begin{equation} \label{minprob}
-S(Q, P) =\frac{\pi}{2\tau_2} |Q^2+\tau P ^2|^2-\ln( h^{(k+2)} (\tau))-
{\ln}(h^{(k +2)}(-\bar{\tau}) -(k+2)\ln(2\tau_2) .
\end{equation}
with respect to $\tau_1, \tau_2$. The  minimum lies at 
\begin{equation}
\tau_1=\frac{Q\cdot P}{P^2}, \qquad 
\tau_2= \frac{1}{P^2}\sqrt {Q^2 P^2-(Q\cdot P)^2}
\end{equation}
Substituting the above values for $\tau_1, \tau_2$ results  in the 
asymptotic behaviour of the index $-B_6$.  The list of the $\Gamma_0(N)$ modular
forms for the models constructed in this paper is provided in table \ref{t6}. 
\begin{table}[H]
\renewcommand{\arraystretch}{0.5}
\begin{center}
\vspace{0.5cm}
\begin{tabular}{|c|c|}
\hline
 & \\
Conjugacy Class  & $h^{(k+2)} (\rho)$ \\ \hline
 &   \\
$pA$ & $\eta^{k+2}(\rho)\eta^{k+2}(p\rho)$\\
 & \\
4B & $\eta^4(4\rho)\eta^2(2\rho)\eta^{4}(\rho)$\\
 & \\
6A & $\eta^2(\rho)\eta^2(2\rho)\eta^2(3\rho)\eta^2(6\rho)$\\
 & \\
8A & $\eta^2(\rho)\eta(2\rho)\eta(4\rho)\eta^2(8\rho)$\\
 & \\
14A &  $\eta(\rho)\eta(2\rho)\eta(7\rho)\eta(14\rho)$ \\
 & \\
15A &  $\eta(\rho)\eta(3\rho)\eta(5\rho)\eta(15\rho)$\\
  & \\
\hline
 & \\
2B & {\Large{$\frac{\eta^8(4\rho)}{\eta^4(2\rho)}$}} \\
 & \\
3B & {\Large{ $\frac{\eta^3(9\rho)}{\eta(3\rho)}$}}\\
 & \\
\hline
\end{tabular}
\end{center}
\vspace{-0.5cm}
\caption{Factorization of $\hat \Phi_k(\rho, \sigma, v) $ as $\lim v\rightarrow 0$ as 
shown in (\ref{hatphifac}), $p\in \{1,2,3,5,7,11,23\}$} \label{t6}
\renewcommand{\arraystretch}{0.5}
\end{table}

Let us now compare this to the behaviour of the 
entropy of single centered large charge $1/4$ BPS dyons in 
these ${\cal N}=4$ theories obtained compactifying 
type II theory on $(K3\times T^2)/\mathbb{Z}_N$ 
where $\mathbb{Z}_N$ acts as the automorphisms 
on $K3$ together with a $1/N$ shift on one of the 
circles of $T^2$. 
Apart from the usual $2$ derivative terms in the 
effective action,  
a one loop computation shows that the coefficient 
of the Gauss-Bonnet term is given by 
\begin{equation}
 \Delta {\cal L} =\phi(a,S) (R_{\mu\nu\rho\sigma}R^{\mu\nu\rho\sigma}-4R_{\mu\nu}R^{\mu\nu}+R^2).
\end{equation}
where  $a, S$ is the axion and dilaton moduli in the heterotic frame. 
The function $\phi(a, S)$ is given by 
\begin{equation}
\phi(a,S)=-\frac{1}{64\pi^2}((k+2)\ln S+\ln h^{(k+2)} (a+iS)+\ln h^{(k+2)} (-a+iS)).
\end{equation}
It is important to note that the $\Gamma_0(N)$ modular form  $h^{(k+2)}(\tau)$ 
for each of the compactifications is identical to the  $\Gamma_0(N)$ form 
that occurs in the factorization (\ref{hatphifac}) \cite{David:2006ud}. 
Now evaluating the 
Hawking-Bekenstein-Wald entropy including the correction due to the 
Gauss-Bonnet term using the entropy function formalism leads to the 
following minimizing problem. 
The entropy is given by  minimizing 
the function
\begin{equation}
{\cal E}(Q, P)=\frac{\pi}{2\tau_2}|Q ^2+\tau P ^2|^2-
\ln h^{(k+2)}(\tau)-\ln h^{(k+2)} (-\tau)-(k+2)\ln (2\tau_2) .
\end{equation}
Here $\tau =  a + i S$. 
The entropy function is identical to  the statistical entropy function  (\ref{minprob})  
which occurred while 
obtaining the asymptotic behaviour of $-B_6$.  
Thus  the partition function $1/\tilde \Phi(\rho, \sigma, v)$ captures 
the degeneracy of 
 large charge single centered $1/4$ BPS black holes in these class of 
 ${\cal N}=4$ compactifications including the correction from the Gauss-Bonnet term.

 The construction of the Siegel modular form given the coefficient of the twisted 
 elliptic genus of $K3$ is quite straight forward and for cyclic orbifolds,  this was first 
 given in \cite{David:2006ud}. 
 Recently the references \cite{Persson:2013xpa,Persson:2015jka,Paquette:2017gmb} 
 extend it for non-cyclic oribolds. 
 It is important to emphasize the there are $2$  modular 
forms associated with the twisted elliptic genus of $K3$. The $\tilde  \Phi_k$ and the $\hat \Phi_k$ 
are  constructed in equations (\ref{siegform}) and (\ref{phihat}) respectively.   
The Fourier expansion of the inverse of $\tilde \Phi_k$  
capture the degeneracy of the $1/4$ BPS dyon and its zeros at $v\rightarrow 0$  are associated with 
the walls of marginal stability of the dyon. The zero's of  $\hat\Phi_k$ are however associated with
the asymptotic growth of the degeneracies for large charges. 
We mention that $\hat \Phi_k$ has not been constructed for the orbifolds listed in this paper 
in the references \cite{Persson:2013xpa,Persson:2015jka,Paquette:2017gmb}. 
We emphasize that our objective in constructing the Siegel  modular 
form $\tilde  \Phi_k$  in particular is to verify that the Fourier expansions of the inverse 
of these forms are integers and positive as predicted by the conjecture of \cite{Sen:2010mz}. 
This was verified earlier by \cite{Sen:2010mz} for the Siegel modular forms $\tilde  \Phi_k$,
associated with the  $pA, p = 1, 2, 3, 5, 7$
orbifolds.  In this next section we 
extend this observation for all the orbifolds discussed in this paper.
To our knowledge, this observation has not been seen in the works of 
\cite{Persson:2013xpa,Persson:2015jka,Paquette:2017gmb}.  
We also emphasize that to obtain this observation the explicit  construction 
of the twisted elliptic genus in all its sectors together with the  normalizations as discussed earlier 
is important.

\subsection{Positivity and integrality of the $1/4$ BPS index}
\label{b66}

In this subsection we provide the list for the index $-B_6$ for low lying charges 
for all the Siegel modular forms  $\tilde \Phi_k$ associated with the 
twisted elliptic genera constructed. 
From the expansion of $\tilde \Phi_k$ in Fourier coefficients in the domain 
(\ref{domain}) together with the expression for $-B_6$ in (\ref{indexb6}) we see that 
the electric charge $Q^2$ is quantized in units of  $2/\mathbb{Z}_N$, while the 
magnetic charge $P^2$ is quantized in units of $2 \mathbb{Z}$ and 
the angular momentum $Q\cdot P$ is an integer. 
We see that the index $-B_6$ for the low lying charges examined 
is always an integer. Furthermore for charges satisfying the condition 
(\ref{pcond}) it is positive. This property is a sufficient condition which
ensures  that single centered black holes  carry zero angular 
momentum. 
One important point to emphasize is that it is possible to obtain 
the Fourier expansion of the Siegel modular forms  for 
low lying charges only after the explicit construction of the twisted elliptic genus. 
As a check on  our Mathematica routines to obtain these Fourier coefficients,
we have verified that our routine 
reproduces all the tables given in \cite{Sen:2010mz} 
for the  $pA$  orbifold of $K3$ with $p=1, 2, 3, 5, 7$.

\begin{table}[H]
\renewcommand{\arraystretch}{0.5}
\begin{center}
\vspace{0.5cm}
\begin{tabular}{|c|c|c|c|c|c|}
\hline
 & & & & & \\
$(Q^2,\;P^2)\;\;\;$ \textbackslash $Q\cdot P$ & -2 & 0 & 1 & 2 & 3\\ 
 & & & & & \\
\hline
 & & & & & \\
(1/2, 2) & -512 & 176 & 8 & 0 & 0\\
(1/2, 4) & -1536 & 896 & 80 & 0 & 0\\ 
(1/2, 6) & -4544 & 3616  & 480 & 0 & 0\\
(1/2, 8) & 11752 & 12848  & 2176 & 24 & 0\\
(1,4) & -4592  & 5024 & 832 & 16 & 0  \\
(1,6) &-13408 &  22464 & 36786  & 224 & 0 \\
(1,8) & -33568 & 88320 & 26176 & 1760 & 0 \\
(3/2, 6) & -37330  & 112316  & 36786  & 2998 & 38\\
(3/2, 8) & -80896 & 491920 & 196960  & 23616  & 592 \\
 & & & & & \\
\hline
\end{tabular}
\end{center}
\vspace{-0.5cm}
\caption{Some results for the index $-B_6$ for  the $4B$
 orbifold of $K3$ for different values of $Q^2$, $P^2$ and $Q\cdot P$ }
\renewcommand{\arraystretch}{0.5}
\end{table}

\begin{table}[H]
\renewcommand{\arraystretch}{0.5}
\begin{center}
\vspace{-0.5cm}
\begin{tabular}{|c|c|c|c|c|c|}
\hline
 & & & & & \\
$(Q^2,\;P^2)\;\;\;$ \textbackslash $Q\cdot P$ & -2 & 0 & 1 & 2 & 3\\ 
 & & & & & \\
\hline
 & & & & & \\
(1/3, 2) & -98 & 40 & 1 & 0 & 0\\
(1/3, 4) & -224 & 148 & 12 & 0 & 0\\ 
(1/3, 6) & -546 & 478 & 49 & 0 & 0\\
(1/3,8) & -1120 & 1352 & 186 & 0 & 0\\
(2/3, 4) & -512 & 592 & 92 & 0 & 0 \\
(2/3, 6) & -1240 & 2080 & 436 & 8 & 0\\
(2/3, 8) & -2504 & 6416 & 1676 & 0 & 0\\
(1, 6) & -2926 & 7880 & 2172 & 116 & 0\\
(1, 10) & -2450 & 81380 & 32300 & 3494 & 49 \\
(1, 12) & -4696 & 234900 & 104176  & 13856 & 316 \\
 & & & & & \\
\hline
\end{tabular}
\end{center}
\vspace{-0.5cm}
\caption{Some results for the index
$-B_6$  for the $6A$ orbifold of $K3$ for different values of $Q^2$, $P^2$ and $Q\cdot P$}
\renewcommand{\arraystretch}{0.5}
\end{table}

\begin{table}[H]
\renewcommand{\arraystretch}{0.5}
\begin{center}
\vspace{-0.5cm}
\begin{tabular}{|c|c|c|c|c|c|}
\hline
 & & & & & \\
$(Q^2,\;P^2)\;\;\;$ \textbackslash $Q\cdot P$ & -2 & 0 & 1 & 2 & 3\\ 
 & & & & & \\
\hline
 & & & & & \\
(1/4, 2) & -60 & 20 & 0 & 0 & 0\\
(1/4, 4) & -120 & 68 & 2 & 0 & 0\\ 
(1/4, 6) & -280 & 196 & 10 & 0 & 0\\
(1/4, 8) & -520 & 504 & 40 & 0 & 0\\
(1/2, 6) &-560 & 724 & 96 & 0 & 0\\
(1/2,8) & -1038 & 1998 & 352 & 2 & 0\\
(3/4, 6) & -1114 & 2280 & 450 & 6 & 0\\
(3/4, 8) & -2024 & 6704 & 1728 & 56 & 0 \\
(3/4, 10) & -3860 &18256  & 5564 & 300 & 0 \\
(3/4, 12)& -6168 & 46456 & 16296  & 1192 & 4 \\
 & & & &  & \\
\hline
\end{tabular}
\end{center}
\vspace{-0.5cm}
\caption{Some results for the index
$-B_6$ for  the $8A$ orbifold of $K3$ for different values of $Q^2$, $P^2$ and $Q\cdot P$}
\renewcommand{\arraystretch}{0.5}
\end{table}

\begin{table}[H]
\renewcommand{\arraystretch}{0.5}
\begin{center}
\vspace{-0.5cm}
\begin{tabular}{|c|c|c|c|c|c|}
\hline
 & & & & & \\
$(Q^2,\;P^2)\;\;\;$ \textbackslash $Q\cdot P$ & -2 & 0 & 1 & 2 & 3\\ 
 & & & & & \\
\hline
 & & & & & \\
(2/11, 2) &-50 & 10 & 0 & 0 & 0\\
(2/11, 4) &-100 & 30 & 0 & 0 & 0\\ 
(2/11, 6) &-200 & 82 & 1 & 0 & 0\\
(4/11, 6) & -400 & 276 & 18 & 0 & 0\\
(6/11, 6)& -800 & 806 & 83 & 0 & 0\\
(6/11, 8) & -1438 & 2064 & 314 & 2 & 0 \\
(6/11, 10) & -2584 & 4962 & 937 & 16 & 0 \\
(6/11, 12) & -4328 & 11132 & 2558  & 72 & 0 \\
(6/11, 22) &-34000& 366378 & 139955 & 12760 & 114 \\
 & & & & & \\
\hline
\end{tabular}
\end{center}
\vspace{-0.5cm}
\caption{Some results for the index $-��B_6$ for the 
$11A$ orbifold of $K3$ for different values of $Q^2$, $P^2$ and $Q\cdot P$}
\renewcommand{\arraystretch}{0.5}
\end{table}

\begin{table}[H]
\renewcommand{\arraystretch}{0.5}
\begin{center}
\vspace{-0.5cm}
\begin{tabular}{|c|c|c|c|c|c|}
\hline
 & & & & &  \\
$(Q^2,\;P^2)\;\;\;$ \textbackslash $Q\cdot P$ &-2 & 0 & 1 & 2 & 3\\ 
 & & & & & \\
\hline
 & & & & & \\
(1/7, 2) & -18 & 4 & 0 & 0 & 0 \\
(1/7, 4) & -24 & 10 & 0 & 0 & 0 \\
(1/7, 6) &-54 & 24 & 0 & 0 & 0 \\
(2/7, 6)& -72 & 70 & 5 & 0 & 0 \\
(2/7, 8) & -96 & 156 & 16 & 0 & 0  \\
(3/7, 8) & -216 & 406 & 65 & 0 & 0 \\
(3/7, 10) &-412 & 890 & 165 & 2 & 0\\
(4/7, 12) &-710 & 4682 & 1443 & 58 & 0 \\
(5/7, 12) &-1180 & 11512 & 4156 & 292 & 0 \\
(5/7, 14) &-1622 & 24744 & 9816 & 908 & 5 \\
 & & & & &  \\
\hline
\end{tabular}
\end{center}
\vspace{-0.5cm}
\caption{Some results the index 
 $-��B_6$  for  the $14A$ orbifold of $K3$ for different values of $Q^2$, $P^2$ and $Q\cdot P$}
\renewcommand{\arraystretch}{0.5}
\end{table}

\begin{table}[H]
\renewcommand{\arraystretch}{0.5}
\begin{center}
\vspace{-0.5cm}
\begin{tabular}{|c|c|c|c|c|c|}
\hline
 & & & & &  \\
$(Q^2,\;P^2)\;\;\;$ \textbackslash $Q\cdot P$ &-2 & 0 & 1 & 2 & 3 \\ 
 & & & & &  \\
\hline
 & & & & & \\
(2/15, 2) & -8 & 4 & 0 & 0 & 0 \\
(2/15, 4) & -16 & 8 & 0 & 0 & 0 \\
(2/15, 6) &-24 & 20 & 0 & 0 & 0\\
(2/5, 8) &-120 & 274 & 45 & 0 & 0 \\
(2/5, 10) &-203 & 578 & 113 & 1 & 0\\
(4/15, 6) &-48 & 50 & 4 & 0 & 0 \\
(4/15, 8) & -80 & 102 & 13 & 0 & 0 \\
(8/15, 12) &-440 & 2844 & 898 & 40 & 0\\
(2/3, 12) & -638 & 6818 & 2498 & 178 & 0\\
(4/5, 18) & 8236 &141252 & 73651 &  12124 & 419 \\
 & & & & &  \\
\hline
\end{tabular}
\end{center}
\vspace{-0.5cm}
\caption{Some results for the index 
$-B_6$  for  the $15A$ orbifold of $K3$ for different values of $Q^2$, $P^2$ and $Q\cdot P$}
\renewcommand{\arraystretch}{0.5}
\end{table}

\begin{table}[H]
\renewcommand{\arraystretch}{0.5}
\begin{center}
\vspace{-0.5cm}
\begin{tabular}{|c|c|c|c|c|c|}
\hline
 & & & & & \\
$(Q^2,\;P^2)\;\;\;$ \textbackslash $Q\cdot P$ &-2 & 0 & 1 & 2 & 3\\ 
 & & & & & \\
\hline
 & & & & & \\
(2/23, 2) & -8 & 1 & 0 & 0 & 0\\
(2/23, 4) & -12 & 3 & 0 & 0 & 0\\ 
(2/23, 6) & -20 & 7 & 1 & 0 & 0\\
(4/23, 6) & -30 & 53 & 6 & 0 & 0\\
(4/23, 8) & -42 & 91 & 11 & 0 & 0\\
(6/23, 6) &-48 & 103 & 23 & 2 & 0\\
(6/23, 8) &-66& 190 & 47 & 4 & 0 \\
(6/23, 10) &-104 & 312 & 74 & 6 & 0 \\
 & & & & & \\
\hline
\end{tabular}
\end{center}
\vspace{-0.5cm}
\caption{Some results for the index $-��B_6$
 for the $23A$ orbifold of $K3$ for different values of $Q^2$, $P^2$ and $Q\cdot P$}
\renewcommand{\arraystretch}{0.5}
\end{table}

\begin{table}[H]
\renewcommand{\arraystretch}{0.5}
\begin{center}
\vspace{-0.5cm}
\begin{tabular}{|c|c|c|c|c|c|}
\hline
 & & & & & \\
$(Q^2,\;P^2)\;\;\;$ \textbackslash $Q\cdot P$ &-2 & 0 & 1 & 2 & 3\\ 
 & & & & & \\
\hline
 & & & & & \\
(1/2, 2) & 320 & 288 & 24 & 0 & 0\\
(1/2, 4) & 0 & 512 & 256 & 0 & 0\\ 
(1/2, 6) & -752 & 1120  & 888 & 48 & 0\\
(1/2, 8) & 384 & 3328  & 2048 & 384 & 0\\
(1,4) & 32  & 4416 & 2240 & 32 & 0  \\
(1,6) & -2304 &  22464 & 13248  & 224 & 0 \\
(1,8) & 5920 & 42944  & 27328 & 5920 & 64  \\
(3/2, 6) & -2008  & 102380   & 66172  & 9032  & 28\\
(3/2, 8) & 59392 & 372736  & 243712  & 59392  & 2048 \\
 & & & & & \\
\hline
\end{tabular}
\end{center}
\vspace{-0.5cm}
\caption{Some results for the index
 $B_6$ for the $2B$ orbifold of $K3$ for different values of $Q^2$, $P^2$ and $Q\cdot P$}
\renewcommand{\arraystretch}{0.5}
\end{table}

\begin{table}[H]
\renewcommand{\arraystretch}{0.5}
\begin{center}
\vspace{-0.5cm}
\begin{tabular}{|c|c|c|c|c|c|}
\hline
 & & & & & \\
$(Q^2,\;P^2)\;\;\;$ \textbackslash $Q\cdot P$ &-2 & 0 & 1 & 2 & 3\\ 
 & & & & & \\
\hline
 & & & & & \\
(2/9, 2) & 0 & 18 & 0 & 0 & 0\\
(2/9, 4) & 18 & 27 & 0 & 0 & 0\\ 
(2/9, 6) & 0 & 78 & 21 & 0 & 0\\
(4/9, 4) & 42 & 150 & 33 & 0 & 0 \\
(4/9, 6) & 0 & 270 & 81 & 0 & 0\\
(4/9, 8) & 0 & 378 & 162 & 0 & 0\\
(2/3, 6) & 0 & 918 & 297 & 0 & 0\\
(2/3, 8) & 0 & 2460 & 1239 & 93 & 0\\
 & & & & & \\
\hline
\end{tabular}
\end{center}
\vspace{-0.5cm}
\caption{Some results for the index $-��B_6$ 
for the $3B$ orbifold of $K3$ for different values of $Q^2$, $P^2$ and $Q\cdot P$}
\renewcommand{\arraystretch}{0.5}
\end{table}

It is interesting to note that the non-geometric orbifolds $11A, 23A, 23B, 2B, 3B$ also 
satisfy the positivity constraints conjectured by 
\cite{Sen:2010mz}.  We have attached the mathematica files which generate the 
Fourier coefficients for the $11A$ and $3B$ orbifolds as ancillary files.

\section{Conclusions}

We have constructed the twisted elliptic genera for $K3$ orbifolded 
by automorphisms corresponding to all the conjugacy classes which 
lie $M_{23}\subset M_{24}$ as well as $2$ conjugacy classes
which does not lie in $M_{23}$. 
Our method involves  the use of the modular transformation property 
of the twisted elliptic genus and discovering identities satisfied by 
$\Gamma_0(N)$ modular forms which relate expansions in $e^{-2\pi i /\tau}$ to 
expansions in $e^{2\pi i \tau}$. 
We also used inputs from $M_{23}$ symmetry to determine
the twisted elliptic genus in sectors which form  sub-orbits under $SL(2, \mathbb{Z})$. 

We then  constructed Siegel modular forms associated with the twisted elliptic genera
that capture the degeneracy of $1/4$ BPS states in 
${\cal N}=4$ theories obtained by compactifying type II theory on 
$(K3\times T^2)/\mathbb{Z}_N$ where $\mathbb{Z}_N$ acts as
a order $N$ automorphism  associated with the 
conjugacy class of $M_{24}$ on $K3$ together with a $1/N$ shift on 
one of the circles of $T^2$. 
We show  that the dyon partition function satisfied the required properties
expected from black hole physics. In particular the Fourier coefficients  of the 
$1/4$ BPS index  are integers and 
certain low lying charges are positive in agreement 
with the conjecture of \cite{Sen:2010mz}.  This is a sufficient
condition 
predicted from the fact   that single centered black holes  carry zero angular 
momentum. 
The construction of the twisted elliptic genus as well as the dyon partition 
function associated with the $4A, 6A, 8A$ classes done in this paper, along with the 
earlier studied cases of $pA$ with $p= 2, 3, 5, 7$ completes this analysis
for all the CHL models. 

It is worthwhile to complete this analysis of this paper for the remaining $9$ conjugacy classes 
of table \ref{t2}.
The construction for the twisted elliptic genera corresponding to these classes 
would required new ingredients. One possible direction is to use positivity and integrality
of the low lying coefficients in the associated Siegel modular form 
to determine the twisted elliptic genera in  the  sectors  which form  sub-orbits
under $SL(2, \mathbb{Z})$.  These conjugacy classes have more than 
one sub-orbits.  One can also verify if the Siegel modular forms 
constructed from  the twisted elliptic genera for these classes provided in the ancillary 
files associated with \cite{Gaberdiel:2012gf} is in agreement with the positivity conjecture of 
\cite{Sen:2010mz}.  

The references \cite{Persson:2013xpa,Persson:2015jka,Paquette:2017gmb} 
has  studied more general  non-cyclic 
 twisted twining elliptic genera of $K3$ 
 than considered in this paper.  
  It  is important to check if the 
 more general twining elliptic genera considered in  these references   admit 
 a $1/4$ BPS dyon partition function with integral Fourier coefficients 
 and obey the  positivity constraints as expected from black hole physics. 
  Recently multiplicative lifts of more general weak Jacobi forms \footnote
 {These were Jacobi forms of weight $0$ but index $>1$.}
 as well as the the Siegel modular forms of $Sp(2, \mathbb{Z})$ of weight
 $35$ and $12$ were studied and were shown to have properties 
 which make them candidates for partition of black holes \cite{Belin:2016knb}. 
 It will be interesting to check if the Fourier coefficients of these Siegel 
 modular forms also satisfy the positivity constraints required from 
 black hole physics.

The discovery of the Mathieu moonshine symmetry 
has provided  useful insights in string  compactifications 
\cite{Kachru,Datta:2015hza,Chattopadhyaya:2016xpa}
as well as provided  new examples  where  precision microscopic counting of 
black holes is possible as seen in this paper. It is certainly worthwhile to explore 
the implication of this symmetry further. 

\acknowledgments

We thank Suresh Govindarajan and Samir Murthy for useful discussions. 
We thank Ashoke Sen for useful discussions, insights, providing 
helpful references  and encouragement 
at various stages of this project.  We thank Mathias Gaberdiel for correspondence
which  helped us compare 
the result of this work with that of  \cite{Gaberdiel:2012gf}.
 A.C 
thanks the Council of Scientific and Industrial Research (CSIR) for funding this project.

\appendix
\section{S-transformations for the $\eta$ function and ${\cal E}_N$ } \label{geneta}

In this appendix we derive identities which for $\eta$ functions and ${\cal E}_N$. 
These identities relate expansions in $e^{-2\pi i /\tau}$ on one side to 
expansions in $e^{2\pi i \tau}$.  
These identities are used in the explicit construction of the twisted elliptic genus 
in all the sectors. 

We begin with the relation between $\eta(-1/\tau + 1/N)$ to 
$\eta ( (\tau -N)/N^2) $
\bea \label{etan1}
\eta\left(\frac{-1}{\tau}+\frac{1}{N}\right)&=&\eta\left(\frac{\tau-N}{N\tau}\right)\\ \nn
&=& \eta\left(\frac{-N\tau}{\tau-N}\right) \left(\frac{(iN\tau)}{\tau-N}\right)^{1/2}\\ \nn
&=&\eta \left(-N-\frac{N^2}{\tau-N}\right) \left(\frac{(iN\tau)}{\tau-N}\right)^{1/2}\\ \nn
&=& e^{-i\pi N/12}\eta \left(\frac{\tau-N}{N^2}\right) \left(\frac{iN\tau}{\tau-N}\right)^{1/2} \left(\frac{-i(\tau-N)}{N^2}\right)^{1/2}\\ \nn
&=& e^{-i\pi N/12}\sqrt {\tau/N}\, \eta \left(\frac{\tau-N}{N^2} \right).
\eea
Now for odd $N$  we find an identity for  $\eta(-1/\tau + 2/N)$
\bea
\eta\left(\frac{-1}{\tau}+\frac{2}{N}\right)&=&\eta\left(\frac{2\tau-N}{N\tau}\right)\\ \nn
&=& \eta\left(\frac{-N\tau}{2\tau-N}\right) \left(\frac{(iN\tau)}{2\tau-N}\right)^{1/2}\\ \nn
&=&\eta \left(-\frac{N}{2}-\frac{N^2}{4\tau-2N}\right) \left(\frac{(iN\tau)}{2\tau-N}\right)^{1/2}.
\eea
Let  $N=2m-1$.
\bea \label{etan2}
\eta \left(-\frac{N}{2}-\frac{N^2}{4\tau-2N}\right) \left(\frac{(iN\tau)}{2\tau-N}\right)^{1/2}&=&\eta \left(\frac{1}{2}-m-\frac{N^2}{4\tau-2N}\right) \left(\frac{(iN\tau)}{2\tau-N}\right)^{1/2}\\ \nn
&=& e^{\pi i/24-\pi i m/12} \frac{\eta^3 \left(-\frac{N^2}{2\tau-N} \right)}{\eta \left(\frac{-N^2}{4\tau-2}\right) \eta \left(\frac{-2N^2}{2\tau-1}\right)} \left(\frac{(iN\tau)}{2\tau-N}\right)^{1/2}\\ \nn
&=& e^{-i\pi m/12} \eta \left(\frac{2\tau-N}{2N^2}+1/2 \right)\sqrt {\tau/N}.
\eea
In the second line of the above equation we have used the identity in (\ref{halfshift1}) to 
relate $\eta$ functions at $1/2$ shifts. 
Then using (\ref{etan1}) and (\ref{etan2}) we obtain 
\begin{equation}\label{etan22}
 \eta\left(\frac{-1}{\tau}+\frac{1}{N}\right) =  
 e^{-i\pi m/12} \eta \left(\frac{2\tau-N}{2N^2}+1/2 \right)\sqrt {\tau/N}, \qquad N = 2m +1.
\end{equation}
Let us now proceed to obtain an identity  involving the shift 
$\eta(-1/\tau - 2/N)$ for odd $N$
\bea\label{etanminus0}
\eta\left(\frac{-1}{\tau}-\frac{2}{N}\right)&=&\eta\left(\frac{-2\tau-N}{N\tau}\right)\\ \nn
&=& \eta\left(\frac{N\tau}{2\tau+N}\right) \left(\frac{(-iN\tau)}{2\tau+N}\right)^{1/2}\\ \nn
&=&\eta \left(\frac{N}{2}-\frac{N^2}{4\tau+2N}\right)  \left(\frac{(-iN\tau)}{2\tau+N}\right)^{1/2}.\\ \nn
\eea
Let $N =2m+1$, then we obtain
\bea \label{etanminus}
\eta \left(\frac{N}{2}-\frac{N^2}{4\tau+2N}\right)  \left(\frac{(-iN\tau)}{2\tau+N}\right)^{1/2}&=&\eta \left(\frac{1}{2}+m-\frac{N^2}{4\tau+2N}\right)  \left(\frac{(-iN\tau)}{2\tau+N}\right)^{1/2}\\ \nn
&=& e^{\pi i/24-\pi i m/12} \frac{\eta^3 \left(-\frac{N^2}{2\tau+N} \right)}{\eta \left(\frac{-N^2}{4\tau+2N}\right) \eta \left(\frac{-2N^2}{2\tau+N}\right)} \left(\frac{(-iN\tau)}{2\tau+N}\right)^{1/2}\\ \nn
&=& -i e^{i\pi m/12} \eta \left(\frac{2\tau+N}{2N^2}+1/2 \right)\sqrt {\tau/N}.
\eea
Here again we have used the identity in n (\ref{halfshift1}) to 
relate $\eta$ functions at $1/2$ shifts. 
Combining (\ref{etanminus0}) and (\ref{etanminus})
we obtain
\begin{equation} \label{etanminus2}
\eta\left(\frac{-1}{\tau}-\frac{2}{N}\right) =
-i e^{i\pi m/12} \eta \left(\frac{2\tau+N}{2N^2}+1/2 \right)\sqrt {\tau/N}, \qquad N = 2m +1.
\end{equation}

Using \ref{etan1}, \ref{etan22}, \ref{etanminus2}  and the 
definition of ${\cal E}_N(\tau)$ we obtain the relations
\bea \label{identenn}
{\cal E}_N(\frac{-1}{N\tau}+\frac{1}{N}) &=& \tau^2 {\cal E}_N(\frac{\tau+N-1}{N}); \\ 
{\cal E}_N(\frac{-1}{N\tau}+\frac{2}{N}) &=& \tau^2 {\cal E}_N(\frac{2\tau+N-1}{2N}), 
\qquad N = 2m +1; \\
{\cal E}_N(\frac{-1}{N\tau}-\frac{2}{N}) &=& \tau^2 {\cal E}_N(\frac{2\tau+N+1}{2N}) 
\qquad N = 2m+1.
\eea
The first relation is true for all $N$ and the last two for $N$ being odd. Therefore 
we can use the last two equations for $N=3, 5, 7$. 
We use these relations repeatedly for 
 obtaining different sectors of  the twisted elliptic genus for the  $14A$ and $15A$ conjugacy class.

Finally using (\ref{halfshift1}) to 
relate $\eta$ functions at $1/2$ shifts we obtain  
\bea
{\cal E}_2(\tau+1/2)&=&-{\cal E}_2(\tau)+2{\cal E}_2(2\tau),\\ \nn
{\cal E}_4(\tau+1/2)&=&\frac{1}{3}(-{\cal E}_2(\tau)+4{\cal E}_2(2\tau)),\\ \nn
{\cal E}_6(\tau+1/2)&=&\frac{2}{5}{\cal E}_2(2\tau)+\frac{4}{5}{\cal E}_3(2\tau)-\frac{1}{5}{\cal E}_2(\tau),\\ \nn
{\cal E}_8(\tau+1/2)&=&\frac{1}{7} \left(-{\cal E}_2(\tau)+2{\cal E}_2(2\tau)+6{\cal E}_4(2\tau)\right), \\ \nn
{\cal E}_{14}(\tau+1/2)&=&\frac{1}{13} \left(-{\cal E}_2(\tau)+2{\cal E}_2(2\tau)+12{\cal E}_7(2\tau)\right), \\ \nn
{\cal E}_{15}(\tau+1/2)&=&-{\cal E}_{15}(\tau)+6{\cal E}_{15}(2\tau)-4{\cal E}_{15}(4\tau) .
\end{eqnarray}
From the definition of ${\cal E}_N$ in terms of the  weight 2 Eisenstein series
\begin{equation}
 {\cal E}_N(\tau)  = \frac{1}{N-1}(N E_2( N\tau) -  E_2(\tau)).
\end{equation}
we obtain the relations 
\begin{eqnarray}
{\cal E}_{15}(\tau+k/3)&=& \frac{1}{7} {\cal E}_3(\tau+\frac{k}{3})+\frac{6}{7} {\cal E}_5 (3\tau), 
\qquad k \in \mathbb{Z}, \\ \nn
{\cal E}_{15}(\tau+k/5)&=& \frac{2}{7} {\cal E}_5(\tau+\frac{k}{3})+\frac{5}{7} {\cal E}_3 (5\tau), 
\qquad k \in \mathbb{Z}.
\end{eqnarray}

\section{Conjugacy class $14A$  and $15A$} \label{frslist}

In this appendix we construct the twisted elliptic genera of $K3$ orbifolded by 
automorphisms corresponding to the conjugacy class $14A$ and $15A$. 

\subsection*{ Conjugacy class $14A$}
\bea
 F^{(0,1)}(\tau, z) = F^{(0,3)}= F^{(0,5)}= F^{(0,9)}= F^{(0,11)}= F^{(0,13)};\\ \nn
=\frac{1}{14}\left[\frac{A}{3}-B \left(-\frac{1}{36}{\cal E}_2(\tau)-\frac{7}{12}{\cal E}_7(\tau)+\frac{91}{36}{\cal E}_{14}(\tau)\right. \right. \\ \nn
\left.\left. -\frac{14}{3} \eta(\tau)\eta(2\tau)\eta(7\tau) \eta(14\tau)\right) \right];
\eea
\bea
F^{(r, rk)}&=&\frac{1}{14}\left[\frac{A}{3}+B \left(-\frac{1}{72}{\cal E}_2(\frac{\tau+k}{2})-\frac{1}{12}{\cal E}_7(\frac{\tau+k}{7})+\frac{13}{72}{\cal E}_{14}(\frac{\tau+k}{14}) \right. \right. \\ \nn
&&\left.\left. -\frac{1}{3} \eta(\tau+k)\eta(\frac{\tau+k}{2})\eta(\frac{\tau+k}{7}) \eta(\frac{\tau+k}{14})
\right) \right];
\eea
where $r$=1,3,5,9,11,13 and $rk$ is Mod 14.

The even twisted sectors with 
odd twining characters can be found by similar manipulations as discussed in detail for the case of the $11A$ 
conjugacy class. This leads to the following equalities. 
\be
F^{(2,13)}=F^{(12,1)}=F^{(6,11)}=F^{(8,3)}=F^{(4,5)}=F^{(10,9)}.
\ee
Combining all these results into a single formula we obtain 
\bea
F^{(2r, 2rk+7)}&=&\frac{1}{14}\left[\frac{A}{3}+B \left(-\frac{1}{6}{\cal E}_2(\tau)-\frac{1}{12}{\cal E}_7(\frac{\tau+k}{7})+\frac{1}{3}{\cal E}_{7}(\frac{2\tau+2k}{7}) \right. \right. \\ \nn
&&\left.\left. -\frac{2}{3} \eta(\tau+k)\eta(2\tau+2k)\eta(\frac{\tau+k}{7}) \eta(\frac{2\tau+2k}{7})\right) \right];
\eea
where $k$ runs from 0 to 6 and except 3 and $r$ from 1 to 6.
Next the following sectors are given by 
\bea
F^{(7,2k+1)}&=&\frac{1}{14}\left[\frac{A}{3}+B \left(-\frac{7}{12}{\cal E}_7(\tau)+\frac{49}{72}{\cal E}_2(\frac{7\tau+1}{2})-\frac{1}{72}{\cal E}_2(\frac{\tau+1}{2}) \right. \right. \\ \nn
&&\left.\left. +\frac{7}{3} 
e^{i\pi 11/12}\eta(\tau)\eta(7\tau)\eta(\frac{\tau+1}{2}) \eta(\frac{7\tau+1}{2})\right) \right];\\ \nn
F^{(7,2k)}&=&\frac{1}{14}\left[\frac{A}{3}+B \left(-\frac{7}{12}{\cal E}_7(\tau)+\frac{49}{72}{\cal E}_2(\frac{7\tau}{2})-\frac{1}{72}{\cal E}_2(\frac{\tau}{2}) \right. \right. \\ \nn
&&\left.\left. +\frac{7}{3} \eta(\tau)\eta(7\tau)\eta(\frac{\tau}{2}) \eta(\frac{7\tau}{2})\right) \right].
\eea
Finally the sectors belonging to the $2A$ and $7A$ sub-orbits are given by 
\bea
F^{(0,0)}&=&\frac{4}{7}A.\\ 
 F^{(0,2k)}&=&\frac{1}{14}\left[A-\frac{7}{4}B{\cal E}_7(\tau)\right] \quad k\; {\rm runs\;from} \;1\;{\rm to}\;6 ,\\ \nn
 F^{(2r,2rk)}&=&\frac{1}{14}\left[A+\frac{1}{4}B{\cal E}_7(\frac{\tau+k}{7})\right]; \quad k\; {\rm runs\;from} \;0\;{\rm to}\;6. \\ 
 F^{(0,7)}&=&\frac{1}{14}\left[\frac{8}{3}A-\frac{4}{3}B{\cal E}_2(\tau)\right] , \\ \nn
 F^{(7,7k)}&=&\frac{1}{14}\left[
 \frac{8}{3}A+\frac{2}{3}B{\cal E}_2(\frac{\tau+k}{2})\right] \quad k\; {\rm runs\;from} \;0\;{\rm to}\;1.
\eea

\subsection*{Conjugacy class $15A$}
\bea
 && F^{(0,1)}(\tau, z) = F^{(0,2)}= F^{(0,4)}= F^{(0,7)}= F^{(0,8)}= F^{(0,11)}= F^{(0,13)}=F^{(0,14)};\\ \nn
&&=\frac{1}{15}\left[
\frac{A}{3}-B \left(-\frac{1}{16}{\cal E}_3(\tau)-\frac{5}{24}{\cal E}_5(\tau)+\frac{35}{16}
{\cal E}_{15}(\tau) -\frac{15}{4} \eta(\tau)\eta(3\tau)\eta(5\tau) \eta(15\tau)\right) \right].
\eea
\bea
F^{(r, rk)}&=&\frac{1}{15}
\left[\frac{A}{3}+B \left(-\frac{1}{48}{\cal E}_3(\frac{\tau+k}{3})-\frac{1}{24}{\cal E}_5(\frac{\tau+k}{5})+\frac{7}{48}{\cal E}_{15}(\frac{\tau+k}{15}) \right. \right. \\ \nn
&&\left.\left. -\frac{1}{4} \eta(\tau+k)\eta(\frac{\tau+k}{3})\eta(\frac{\tau+k}{5}) 
\eta(\frac{\tau+k}{15})\right) \right];
\eea
where $r$=1,2,4,7,8,11,13,14 and $rk$ is mod 15.
The sectors belonging to the $5A$ and $3A$ sub-orbits are given by 
\bea
F^{(0,0)}&=&\frac{8}{15}A.\\ 
 F^{(0,3k)}&=&\frac{1}{15}\left(\frac{4}{3}A-\frac{5}{3}B{\cal E}_5(\tau)\right) \quad k\; {\rm runs\;from} \;1\;{\rm to}\;4;\\ \nn
 F^{(3r,3rk)}&=&\frac{1}{15}\left(\frac{4}{3}A+\frac{1}{3}B{\cal E}_5(\frac{\tau+k}{5})\right); \quad k\; {\rm runs\;from} \;0\;{\rm to}\;4. \\ 
 F^{(0,5k)}&=&\frac{1}{15}\left(2A-\frac{3}{2}B{\cal E}_3(\tau)\right) ;\\ \nn
 F^{(5r,5rk)}&=&\frac{1}{15}\left(2A+\frac{1}{2}B{\cal E}_3(\frac{\tau+k}{3})\right) \quad k\; {\rm runs\;from} \;0\;{\rm to}\;2.
\eea
Finally the remaining sectors are given by 
\bea
F^{(3r, 5s+3rk)}&=&\frac{1}{15}\left(\frac{A}{3}+B \left(-\frac{1}{4}{\cal E}_3(\tau)-\frac{1}{24}{\cal E}_5(\frac{\tau+k}{5})+\frac{3}{8}{\cal E}_{5}(\frac{3\tau+3k}{5}) \right. \right. \\ \nn
&&\left.\left. -\frac{3}{4} \eta(\tau+k)\eta(3\tau+3k)\eta(\frac{\tau+k}{5}) \eta(\frac{3\tau+3k}{5})\right) \right);
\eea
where $k$ runs from 0 to 4 and $s$=1 to 4.
\bea
F^{(5r, 3s+5rk)}&=&\frac{1}{15}\left(\frac{A}{3}+B \left(\frac{5}{24}{\cal E}_5(\tau)+\frac{1}{12}{\cal E}_3(\frac{\tau+k}{3})-\frac{5}{24}{\cal E}_{5}(\frac{\tau+k}{3}) \right. \right. \\ \nn
&&\left.\left. +\frac{5}{4} \eta(\tau+k)\eta(5\tau+5k)\eta(\frac{\tau+k}{3}) \eta(\frac{5\tau+5k}{3})\right) \right);
\eea
where $k$ runs from 0 to 2 and $s$=1 to 2.

The low lying coefficients of the
twisted elliptic genus in conjugacy classes $14A$ as well as $15A$ satisfy
\begin{eqnarray}
c^{(0, s)}(\pm 1) &=& \frac{2}{N}, \qquad 
\sum_{s=0}^{N-1} c^{(0, s)} (\pm 1)  = 2, \\ \nonumber
\sum_{s=0}^{N-1} c^{(0, )} (0) &=& 0, \qquad 
 \sum_{r, s=0}^{N-1} F^{(r, s)}(\tau, z) = 8 A(\tau, z) 
\end{eqnarray}


\begin{thebibliography}{10}

\bibitem{Dijkgraaf:1996it}
R.~Dijkgraaf, E.~P. Verlinde, and H.~L. Verlinde, {\it {Counting dyons in N=4
  string theory}},  {\em Nucl. Phys.} {\bf B484} (1997) 543--561,
  [\href{http://arxiv.org/abs/hep-th/9607026}{{\tt hep-th/9607026}}].

\bibitem{LopesCardoso:2004law}
G.~Lopes~Cardoso, B.~de~Wit, J.~Kappeli, and T.~Mohaupt, {\it {Asymptotic
  degeneracy of dyonic N = 4 string states and black hole entropy}},  {\em
  JHEP} {\bf 12} (2004) 075, [\href{http://arxiv.org/abs/hep-th/0412287}{{\tt
  hep-th/0412287}}].

\bibitem{Jatkar:2005bh}
D.~P. Jatkar and A.~Sen, {\it {Dyon spectrum in CHL models}},  {\em JHEP} {\bf
  04} (2006) 018, [\href{http://arxiv.org/abs/hep-th/0510147}{{\tt
  hep-th/0510147}}].

\bibitem{David:2006yn}
J.~R. David and A.~Sen, {\it {CHL Dyons and Statistical Entropy Function from
  D1-D5 System}},  {\em JHEP} {\bf 11} (2006) 072,
  [\href{http://arxiv.org/abs/hep-th/0605210}{{\tt hep-th/0605210}}].

\bibitem{Shih:2005uc}
D.~Shih, A.~Strominger, and X.~Yin, {\it {Recounting Dyons in N=4 string
  theory}},  {\em JHEP} {\bf 10} (2006) 087,
  [\href{http://arxiv.org/abs/hep-th/0505094}{{\tt hep-th/0505094}}].

\bibitem{Chaudhuri:1995fk}
S.~Chaudhuri, G.~Hockney, and J.~D. Lykken, {\it {Maximally supersymmetric
  string theories in {$D < 10$} }},  {\em Phys. Rev. Lett.} {\bf 75} (1995)
  2264--2267, [\href{http://arxiv.org/abs/hep-th/9505054}{{\tt
  hep-th/9505054}}].

\bibitem{David:2006ji}
J.~R. David, D.~P. Jatkar, and A.~Sen, {\it {Product representation of Dyon
  partition function in CHL models}},  {\em JHEP} {\bf 06} (2006) 064,
  [\href{http://arxiv.org/abs/hep-th/0602254}{{\tt hep-th/0602254}}].

\bibitem{Dabholkar:2006xa}
A.~Dabholkar and S.~Nampuri, {\it {Spectrum of dyons and black holes in CHL
  orbifolds using Borcherds lift}},  {\em JHEP} {\bf 11} (2007) 077,
  [\href{http://arxiv.org/abs/hep-th/0603066}{{\tt hep-th/0603066}}].

\bibitem{Nikulin}
V.~V. Nikulin, {\it {Finite automorphism groups of K\"{a}hler {$K3$}
  surfaces}},  {\em Trans. Moscow. Math. Soc.} {\bf 38} (1979) 71.

\bibitem{Chaudhuri:1995dj}
S.~Chaudhuri and D.~A. Lowe, {\it {Type IIA heterotic duals with maximal
  supersymmetry}},  {\em Nucl. Phys.} {\bf B459} (1996) 113--124,
  [\href{http://arxiv.org/abs/hep-th/9508144}{{\tt hep-th/9508144}}].

\bibitem{Aspinwall:1995fw}
P.~S. Aspinwall, {\it {Some relationships between dualities in string theory}},
   {\em Nucl. Phys. Proc. Suppl.} {\bf 46} (1996) 30--38,
  [\href{http://arxiv.org/abs/hep-th/9508154}{{\tt hep-th/9508154}}].

\bibitem{Eguchi:2010ej}
T.~Eguchi, H.~Ooguri, and Y.~Tachikawa, {\it {Notes on the K3 Surface and the
  Mathieu group $M_{24}$}},  {\em Exper. Math.} {\bf 20} (2011) 91--96,
  [\href{http://arxiv.org/abs/1004.0956}{{\tt arXiv:1004.0956}}].

\bibitem{Cheng:2010pq}
M.~C. Cheng, {\it {K3 Surfaces, N=4 Dyons, and the Mathieu Group M24}},  {\em
  Commun.Num.Theor.Phys.} {\bf 4} (2010) 623--658,
  [\href{http://arxiv.org/abs/1005.5415}{{\tt arXiv:1005.5415}}].

\bibitem{Eguchi:2010fg}
T.~Eguchi and K.~Hikami, {\it {Note on twisted elliptic genus of $K3$
  surface}},  {\em Phys. Lett.} {\bf B694} (2011) 446--455,
  [\href{http://arxiv.org/abs/1008.4924}{{\tt arXiv:1008.4924}}].

\bibitem{Gaberdiel:2010ch}
M.~R. Gaberdiel, S.~Hohenegger, and R.~Volpato, {\it {Mathieu twining
  characters for K3}},  {\em JHEP} {\bf 09} (2010) 058,
  [\href{http://arxiv.org/abs/1006.0221}{{\tt arXiv:1006.0221}}].

\bibitem{Sen:2007qy}
A.~Sen, {\it {Black Hole Entropy Function, Attractors and Precision Counting of
  Microstates}},  {\em Gen. Rel. Grav.} {\bf 40} (2008) 2249--2431,
  [\href{http://arxiv.org/abs/0708.1270}{{\tt arXiv:0708.1270}}].

\bibitem{Dabholkar:2012zz}
A.~Dabholkar and S.~Nampuri, {\it {Quantum black holes}},  {\em Lect. Notes
  Phys.} {\bf 851} (2012) 165--232, [\href{http://arxiv.org/abs/1208.4814}{{\tt
  arXiv:1208.4814}}].

\bibitem{Sen:2010mz}
A.~Sen, {\it {How Do Black Holes Predict the Sign of the Fourier Coefficients
  of Siegel Modular Forms?}},  {\em Gen. Rel. Grav.} {\bf 43} (2011)
  2171--2183, [\href{http://arxiv.org/abs/1008.4209}{{\tt arXiv:1008.4209}}].

\bibitem{Gaberdiel:2013psa}
M.~R. Gaberdiel, A.~Taormina, R.~Volpato, and K.~Wendland, {\it {A K3 sigma
  model with $\mathbb{Z}^8_2$ : $\mathbb{M}_{20}$ symmetry}},  {\em JHEP} {\bf
  02} (2014) 022, [\href{http://arxiv.org/abs/1309.4127}{{\tt
  arXiv:1309.4127}}].

\bibitem{Gaberdiel:2012gf}
M.~R. Gaberdiel, D.~Persson, H.~Ronellenfitsch, and R.~Volpato, {\it
  {Generalized Mathieu Moonshine}},  {\em Commun. Num. Theor Phys.} {\bf 07}
  (2013) 145--223, [\href{http://arxiv.org/abs/1211.7074}{{\tt
  arXiv:1211.7074}}].

\bibitem{David:2006ud}
J.~R. David, D.~P. Jatkar, and A.~Sen, {\it {Dyon spectrum in generic N=4
  supersymmetric Z(N) orbifolds}},  {\em JHEP} {\bf 01} (2007) 016,
  [\href{http://arxiv.org/abs/hep-th/0609109}{{\tt hep-th/0609109}}].

\bibitem{Persson:2013xpa}
D.~Persson and R.~Volpato, {\it {Second Quantized Mathieu Moonshine}},  {\em
  Commun. Num. Theor. Phys.} {\bf 08} (2014) 403--509,
  [\href{http://arxiv.org/abs/1312.0622}{{\tt arXiv:1312.0622}}].

\bibitem{Persson:2015jka}
D.~Persson and R.~Volpato, {\it {Fricke S-duality in CHL models}},  {\em JHEP}
  {\bf 12} (2015) 156, [\href{http://arxiv.org/abs/1504.0726}{{\tt
  arXiv:1504.0726}}].

\bibitem{Paquette:2017gmb}
N.~M. Paquette, R.~Volpato, and M.~Zimet, {\it {No More Walls! A Tale of
  Modularity, Symmetry, and Wall Crossing for 1/4 BPS Dyons}},  {\em JHEP} {\bf
  05} (2017) 047, [\href{http://arxiv.org/abs/1702.0509}{{\tt
  arXiv:1702.0509}}].

\bibitem{Bringmann:2012zr}
K.~Bringmann and S.~Murthy, {\it {On the positivity of black hole degeneracies
  in string theory}},  {\em Commun. Num. Theor Phys.} {\bf 07} (2013) 15--56,
  [\href{http://arxiv.org/abs/1208.3476}{{\tt arXiv:1208.3476}}].

  \bibitem{zagier}
C. Itzykson et.al, {\em
  {From Number Theory to Physics}}, Chapter 4 by Don Zagier .
Springer 1992.

\bibitem{link}
\url{http://www.lmfdb.org/ModularForm/GL2/Q/holomorphic/23/2/1/}
  
  
  
\bibitem{Govindarajan:2009qt}
S.~Govindarajan and K.~Gopala~Krishna, {\it {BKM Lie superalgebras from dyon
  spectra in Z(N) CHL orbifolds for composite N}},  {\em JHEP} {\bf 05} (2010)
  014, [\href{http://arxiv.org/abs/0907.1410}{{\tt arXiv:0907.1410}}].

\bibitem{Sen:2007vb}
A.~Sen, {\it {Walls of Marginal Stability and Dyon Spectrum in N=4
  Supersymmetric String Theories}},  {\em JHEP} {\bf 05} (2007) 039,
  [\href{http://arxiv.org/abs/hep-th/0702141}{{\tt hep-th/0702141}}].

\bibitem{Dabholkar:2007vk}
A.~Dabholkar, D.~Gaiotto, and S.~Nampuri, {\it {Comments on the spectrum of CHL
  dyons}},  {\em JHEP} {\bf 01} (2008) 023,
  [\href{http://arxiv.org/abs/hep-th/0702150}{{\tt hep-th/0702150}}].

\bibitem{LopesCardoso:2006ugz}
G.~Lopes~Cardoso, B.~de~Wit, J.~Kappeli, and T.~Mohaupt, {\it {Black hole
  partition functions and duality}},  {\em JHEP} {\bf 03} (2006) 074,
  [\href{http://arxiv.org/abs/hep-th/0601108}{{\tt hep-th/0601108}}].

\bibitem{Belin:2016knb}
A.~Belin, A.~Castro, J.~Gomes, and C.~A. Keller, {\it {Siegel Modular Forms and
  Black Hole Entropy}},  {\em JHEP} {\bf 04} (2017) 057,
  [\href{http://arxiv.org/abs/1611.0458}{{\tt arXiv:1611.0458}}].

\bibitem{Kachru}
M.~C. Cheng, X.~Dong, J.~Duncan, J.~Harvey, S.~Kachru, et~al., {\it {Mathieu
  Moonshine and N=2 String Compactifications}},  {\em JHEP} {\bf 1309} (2013)
  030, [\href{http://arxiv.org/abs/1306.4981}{{\tt arXiv:1306.4981}}].

\bibitem{Datta:2015hza}
S.~Datta, J.~R. David, and D.~Lust, {\it {Heterotic string on the CHL orbifold
  of K3}},  {\em JHEP} {\bf 02} (2016) 056,
  [\href{http://arxiv.org/abs/1510.0542}{{\tt arXiv:1510.0542}}].

\bibitem{Chattopadhyaya:2016xpa}
A.~Chattopadhyaya and J.~R. David, {\it {${\cal N}=2$ heterotic string
  compactifications on orbifolds of $K3\times T^2$}},  {\em JHEP} {\bf 01}
  (2017) 037, [\href{http://arxiv.org/abs/1611.0189}{{\tt arXiv:1611.0189}}].

\end{thebibliography}

\providecommand{\href}[2]{#2}\begingroup\raggedright\endgroup

\end{document}